\documentclass[11pt,a4paper]{amsart}
\usepackage[utf8]{inputenc}
\usepackage[a4paper,margin=2.5cm]{geometry}
\usepackage{amsmath,amssymb,amsthm,mathtools,abraces}
\usepackage{stmaryrd}
\usepackage{color,xcolor}
\usepackage[hypertexnames=false,hyperfootnotes=false,colorlinks=true,linkcolor=blue,%
citecolor=purple,filecolor=magenta,urlcolor=cyan,unicode,pagebackref=true]{hyperref}
\usepackage{yhmath}
\usepackage{verbatim}
\usepackage{enumitem} 
\usepackage{bbm}
\usepackage{csquotes}

\newcommand\be{\begin{equation}}
\newcommand\ee{\end{equation}}

\newcommand\p{\partial}

\newcommand\Tr{{\rm Tr}\,}
\newcommand\diag{{\rm diag}\,}

\def\lvac{\left <0\right |}
\def\rvac{\left |0\right >}
\DeclareMathOperator{\GL}{GL}

\DeclareMathOperator{\gl}{\mathfrak{gl}}

\DeclareMathOperator{\Mat}{Mat}

\def\psistar{\psi^{*}}
\def\lbr{\left <}
\def\rbr{\right >}
\def\normord{ {\scriptstyle {{\bullet}\atop{\bullet}}} }
\def\lvacn{\left <n\right |}
\def\rvacn{\left |n\right >}

\makeatletter
\@namedef{subjclassname@2020}{%
  \textup{2020} Mathematics Subject Classification}
\makeatother

\newtheorem{theorem}{Theorem}
\newtheorem{lemma}{Lemma}[section]
\newtheorem{proposition}[lemma]{Proposition}
\newtheorem{corollary}[lemma]{Corollary}
\newtheorem{remark}{Remark}[section]

\newtheorem*{theorem*}{Theorem}

\theoremstyle{definition}
\newtheorem{definition}{Definition}

\numberwithin{equation}{section}
\usepackage{filecontents}

\title[Matrix models for the nested hypergeometric tau-functions]{
Matrix models for the nested hypergeometric tau-functions}

\author{Alexander Alexandrov}

\address{Center for Geometry and Physics, Institute for Basic Science (IBS), Pohang 37973, Korea
}

\email{ {\tt alexandrovsash at gmail.com}}


\date{\today}

\begin{document}

\begin{abstract} 
We introduce and investigate a family of tau-functions of the 2D Toda hierarchy, 
which is a natural generalization of the hypergeometric family associated with Hurwitz numbers. 
For this family we prove a skew Schur function expansion formula. For arbitrary rational weight generating functions we construct the multi-matrix models. Two different types
of cut-and-join descriptions are derived. Considered examples include generalized fully simple maps, which we identify with the recently introduced skew hypergeometric tau-functions.
\end{abstract}

\maketitle

\tableofcontents



\def\thefootnote{\arabic{footnote}}


\section{Introduction}
\setcounter{equation}{0}

The counting of objects related to Riemann surfaces is often governed by the integrable hierarchies of the KP/Toda type. 
Since the pioneering works of Okounkov and Pandharipande \cite{O,P} we know that simple Hurwitz numbers, which count ramified coverings of the sphere,
generate tau-functions of the 2D Toda hierarchy. A more general family of weighted Hurwitz numbers \cite{GJ2,GPH} corresponds to a hypergeometric (or Orlov--Scherbin) family of 
tau-functions, introduced in \cite{KMMM} and further investigated in \cite{OS}. This family of tau-functions attracts a lot of attention because of its spectacular properties. In particular, many interesting hypergeometric tau-functions can be described by various versions of matrix models and cut-and-join operators \cite{GJ,ZJZ,Orlov,O2,O1,Eynard,MMRP, AMMN,GGPN,AC,Zograf,Orlovchain,Ferm,Adr,WLZZ}.

In this paper, inspired by the recent progress in matrix models and cut-and-join operators on one side \cite{WLZZ,Adr,Oconj,MMCH}, and enumerative geometry on another \cite{BGF,BDKS,BCDG,BCGF}, we introduce a family of tau-functions, which can be considered as a deep generalization of the hypergeometric family. Our definition is based on the free fermion description, inherent to the solitonic integrable hierarchies. We call the constructed tau-functions the {\em nested hypergeometric tau-functions}. For this family we derive the skew Schur function expansion and the cut-and-join description, which are direct generalizations of the constructions for the hypergeometric tau-functions. 

Hypergeometric tau-functions are closely related to matrix models, and one of the main goals of this paper is to generalize this relation to the nested hypergeometric tau-functions. For this purpose we restrict ourself to the tau-functions with diagonal group operators described by the weight generating functions of the form
\be
G(z)=\frac{\prod_{i=1}^\ell(1+u_iz)}{\prod_{j=1}^n(1+v_jz)}e^{wz}.
\ee
This corresponds to an arbitrary combination of the weight generating function associated with the strictly monotone, monotone, and simple Hurwitz numbers,
which makes the family natural for the enumerative geometry and topological string applications.  For this family of the nested hypergeometric tau-functions we provide a general construction of the multi-matrix model. This multi-matrix model is a combination of the normal, unitary, and complex matrix integrals and has a chain structure, which allows us to reduce it to the eigenvalue integral. The eigenvalue integral is a combination of the perturbative Gaussian integrals and elementary residues. 

For more specific combinations of the diagonal group elements the nested hypergeometric tau-functions can also be described by the cut-and-join operators of  \cite{WLZZ,Adr,Oconj,MMCH}. As usual, the tau-functions are described by the action of the exponential operators on the elementary generating function (an exponential of a linear combination of variables, or just a constant), however, for the nested hypergeometric family this action contains several non-commutative exponential operators. 

We consider in detail several examples. For the simplest case of the hypergeometric tau-functions, associated with weighted Hurwitz numbers, we provide an example of the matrix integral for the rational weight generating function and discuss the ambiguity, which helps to simplify the matrix model. We also reconsider different cut-and-join operator descriptions and relations between them. The next family of examples of the nested hypergeometric tau-functions is associated with the skew hypergeometric tau-functions, recently introduces in  \cite{Oconj,MMCH}. We identify a particular representative of this family with the generating function of the fully simple maps (more generally, hypermaps), recently investigated in \cite{BGF,BDKS}. In particular, we prove a skew Schur expansion formula for the generating function of fully simple maps and its generalization. For this family of  tau-functions, which includes the fully simple case, we provide several different cut-and-join descriptions. We discuss the interrelations between different cut-and-join formulas and connection to the hypergeometric case. Some other examples are discussed, but their relation to the enumerative geometry invariants is not clear yet.

Many interesting questions are  beyond the scope of this paper. In particular, Sato Grassmannian description and associated Virasoro type linear constraints for the nested hypergeometric tau-functions are not
discussed here. Fermionic representation should help us to obtain the Kac--Schwarz operators, including the canonical ones, and to derive
linear constraints. Matrix models can also be helpful for this purpose. 

Many elements of our general construction can be immediately applied to other families of the generating functions. Perhaps the most interesting direction is associated with integrable hierarchies of type B. 
For this case, the generalization which includes the definition of the nested hypergeometric tau-function of the 2-component BKP hierarchy, free neutral fermion description, skew Schur Q-function expansion, and cut-and-join operators is straightforward. However, the general construction of the matrix integral can be more tricky.
Deformation to the skew Jack polynomial version and associated $\beta$-deformed matrix integrals, as well as $(q,t)$-deformations
are also very attractive and, for some elements of the general construction, are direct.
We will return to these topics in the upcoming publications.

\bigbreak

\subsection{Organization of the paper} In Section \ref{S2}, we recall the main ingredients of the description of tau-functions in terms of free fermions. We use this free fermion formulation to introduce the 
nested hypergeometric tau-functions in Section \ref{S3}. In this section we also prove the skew Schur expansion formula and cut-and-join recursion for the general nested hypergeometric tau-function. Section
\ref{S4} is devoted to the nested hypergeometric tau-functions with the diagonal group elements of specific form. For all these tau-functions we construct multi-matrix models, which can be reduced to the eigenvalue integrals, 
and for a more specific family we construct another cut-and-join description. In Section \ref{S5} we apply the general construction to the examples and compare them with known matrix models and enumerative geometry invariants. In Section 
\ref{S6} we comment on the partition functions with broken KP integrability.

\bigbreak

\noindent{\bf Acknowledgments.} This work was supported by IBS-R003-D1. The author is grateful to IHES and Haifa University for hospitality. We would like to
thank the referees for useful remarks and suggestions.

\section{Tau-functions and free fermions}\label{S2}

In this section, we recall the description of tau-functions of the Kadomtsev--Petviashvili (KP) hierarchy, modified 
Kadomtsev--Petviashvili (MKP) hierarchy, and 2D Toda lattice hierarchy in terms of free fermions (or, equivalently, infinite wedge space). We also provide some details on the Schur functions
and their relation to the free fermions and tau-functions.
We refer the reader to \cite{Macdonald,JMbook,AZ}  for further details and proofs.

Let us introduce the free fermions  $\psi_n , \psistar_{n}$, $n\in \mathbb{Z}$ satisfying the canonical anticommutation relations
\be\label{anti}
[\psi_n , \psi_m ]_+ = [\psistar_n, \psistar_m]_+=0, \quad
[\psi_n , \psistar_m]_+=\delta_{mn}.
\ee
They generate
an infinite dimensional Clifford algebra. 

Next, we introduce a vacuum state $\left |0\rbr$, which satisfies: 
\be
\psi_n \rvac =0, \quad n< 0; \quad \quad \quad
\psistar_n \rvac =0, \quad n\geq 0.
\ee
Similarly, the dual vacuum state has the properties
\be
\lvac \psistar_n  =0, \quad n< 0; \quad \quad \quad
\lvac \psi_n  =0, \quad n\geq 0.
\ee
With respect to the vacuum $\rvac$, the operators $\psi_n$ with
$n<0$ and $\psistar_n$ with $n\geq 0$ are annihilation operators,
while the operators $\psistar_n$ with $n<0$ and
$\psi_n$ with $n\geq 0$ are creation operators. 
We also introduce ``shifted''  vacua $\rvacn$ and $\lvacn$
defined as   
\begin{align}\label{vacdefr}
\rvacn = \left \{
\begin{array}{l}
\psi_{n-1}\ldots \psi_1 \psi_0 \rvac , \,\,\,\,\, n> 0,
\\ \\
\psistar_n \ldots \psistar_{-2}\psistar_{-1}\rvac , \,\,\,\,\, n<0,
\end{array} \right.
\end{align}
\begin{align}\label{vacdefl}
\lvacn = \left \{
\begin{array}{l}
\lvac \psistar_{0}\psistar_{1}\ldots \psistar_{n-1} , \,\,\,\,\, n> 0,
\\ \\
\lvac \psi_{-1}\psi_{-2}\ldots \psi_{n} , \,\,\,\,\, n<0.
\end{array} \right.
\end{align}
For them we have 
\begin{equation}
\begin{split}
 \psi_m \rvacn &=0, \quad m < n; 
\quad \quad \quad
\psistar_m \rvacn =0, \quad m \ge n, \\
\lvacn  \psi_{m}&=0 , \quad m \ge n; 
\quad \quad \quad
\lvacn  \psistar_{m}=0 , \quad m < n.
\end{split}
\end{equation}

Excited states over the vacuum 
$\rvac$ are obtained by action of the creation operators. We say that the operators $\psi_j$ increase the charge by one, and the operators $\psistar_j$ decrease the charge by one. 
Let us introduce a convenient basis
of states with definite charge in the fermionic Fock space ${\mathcal H}_{F}$.
The basis states $\left |\lambda , n\rbr$ are
parametrized by the charge $n$ and partitions $\lambda$ in the following way.
Given a partition $\lambda =
(\lambda_1 , \ldots , \lambda_{\ell})$ with $\ell =\ell (\lambda )$
nonzero parts, let
$(\vec \alpha |\vec \beta )=(\alpha_1, \ldots , \alpha_{d(\lambda )}|
\beta_1 , \ldots , \beta_{d(\lambda )})$ be the Frobenius notation
for the partition $\lambda$. 
Here $d(\lambda )$ is the number of
boxes in the main diagonal and $\alpha_i =\lambda_i -i$,
$\beta_i =\lambda'_i -i$, where $\lambda'$ is the transposed
(reflected about the main diagonal) diagram $\lambda$. Then
\be\label{lambda1}
\begin{array}{l}
\left |\lambda , n\rbr :=
\psistar_{n-\beta_1 -1}\ldots \psistar_{n-\beta_{d(\lambda )}\! -1}\,
\psi_{n+\alpha_{d(\lambda )}}\ldots \psi_{n+\alpha_1}\rvacn ,
\\ \\
\lbr \lambda , n \right |:=
\lvacn \psistar_{n+\alpha_1}\ldots \psistar_{n+\alpha_{d(\lambda )}}\,
\psi_{n-\beta_{d(\lambda )}\! -1}\ldots \psi_{n-\beta_1 -1} .
\end{array}
\ee
The state $\left |\lambda , n\rbr$ has the charge $n$
with respect to the vacuum state $\rvac$. For the empty diagram
$\left < \emptyset , n\right |=\lvacn$, 
$\left | \emptyset , n\right >=\rvacn$.

Let
\be\label{boflambda}
b(\lambda )=\sum_{i=1}^{d(\lambda )}(\beta_i +1).
\ee
The basis states $\left |\lambda , n \right >$ and $\left <\lambda , n \right |$ can be constructed from a vacuum
in another, equivalent way which is sometimes more convenient
\be
\begin{array}{l}
\left |\lambda , n \right >=(-1)^{b(\lambda )}
\, \psi_{n+\lambda_1 -1}\psi_{n+\lambda_2 -2}\ldots 
\psi_{n+\lambda_{\ell} -\ell}\left | n-\ell \right >,
\\ \\
\left <\lambda , n \right |=(-1)^{b(\lambda )}
\left < n-\ell \right | \psistar_{n+\lambda_{\ell} -\ell}
\ldots \psistar_{n+\lambda_2 -2}\psistar_{n+\lambda_1 -1}.
\end{array}
\ee

The basis vectors (\ref{lambda1})
are orthonormal with respect to the
scalar product induced by the expectation value:
$$
\lbr \lambda , n\right | \left. \mu , m \rbr =
\delta_{mn}\delta_{\lambda \mu}.
$$

The normal ordering $\normord (\ldots )\normord $ with respect
to the vacuum $\rvac$ is defined as 
follows: all annihilation operators
are moved to the right and all creation operators are moved to
the left, taking into account that the factor $(-1)$ appears 
each time two neighboring 
fermionic operators exchange their positions. 
Normally ordered bilinear combinations $X_B=\sum_{mn} B_{mn}\normord\psistar_m \psi_n\normord$
of the fermions, with certain conditions
on the matrix $B = (B_{mn})$, generate an
infinite-dimensional Lie algebra $\gl(\infty)$. Exponentiating these expressions, one obtains
an infinite dimensional group (a version
of $\GL(\infty )$) with the group elements 
\begin{equation}\label{gl}
G=\exp \Bigl (\sum_{i, k \in {\mathbb Z }}B_{ik}\normord\psistar_i \psi_k\normord\Bigr ).
\end{equation}

Let us consider the operators 
\be\label{Jk}
J_k =\sum_{j\in {\mathbb Z}}\normord \psi_j \psistar_{j+k}\normord.
\ee
These operators belong to $\gl(\infty)$ Lie algebra and span the Heisenberg algebra
\be\label{Heis}
\left[J_k, J_l\right]= k \delta_{k+l,0}.
\ee
Operators $J_k$ with positive and negative $k$ annihilate right and left vacua respectively,
\be
J_k\rvacn=\lvacn J_{-k}=0,\quad \quad \quad k>0,
\ee
and for the zeroth mode we have
\be
J_0\rvacn = n\rvacn,\quad \quad \lvacn J_0=n \lvacn.
\ee

For any group element $G$ a tau-function of the MKP hierarchy is given by a vacuum expectation value
\begin{equation}\label{tau}
\tau_n ({\bf t})=\lvacn {\gamma_+ ({\bf t})}G\rvacn,
\end{equation}
where we introduce
\be
\gamma_\pm({\bf t})=e^{\sum_{k>0} t_k J_{\pm k}}.
\ee
It depends on
the variables ${\bf t}=\{t_1, t_2, \ldots \}$, usually called times, and on $n \in \mathbb{Z}$.

The MKP hierarchy relates $\tau_m$ to $\tau_{n}$ for any $m-n \in \mathbb{Z}_{\geq 0}$.
It can be described by the bilinear Hirota identity 
\begin{equation}\label{bi1}
\oint_{{\infty}} z^{m-n} e^{\sum_{k=1}^\infty (t_k-t_k')z^k}
\,\tau_{m} ({\bf t}-[z^{-1}])\,\tau_{n} ({\bf t'}+[z^{-1}])dz =0,
\end{equation}
where we use the standard short-hand notations
\be
{\bf t}\pm [z^{-1}]:= \bigl \{ t_1\pm   
z^{-1}, t_2\pm \frac{1}{2}z^{-2}, 
t_3 \pm \frac{1}{3}z^{-3}, \ldots \bigr \}.
\ee
For a fixed $n$, $\tau_n ({\bf t})$ is a tau-function of the KP hierarchy and satisfies \eqref{bi1} with $m=n$.

More generally, for any group element $G$, the vacuum expectation value
\be\label{stancor}
\tau_n({\bf t},{\bf \tilde t})=\lvacn  \gamma_+({\bf t})\, G\,  \gamma_- ({\bf \tilde t}) \rvacn 
\ee
is a tau-function of the 2D Toda lattice hierarchy. It is also a tau-function of the 2-component MKP
hierarchy.  

Let
\be
\widehat{J}_k =
\left \{ 
\begin{aligned}
&\frac{\p}{\p t_k}\quad \quad \quad & \mathrm{for} \quad k>0,\\[2pt]
&n &\mathrm{for} \quad k=0,\\[2pt]
&-kt_{-k}  &\mathrm{for} \quad k<0,
 \end{aligned}
\right.
\ee
then
\be\label{JtoJ}
\widehat{J}_k \cdot \lvacn \gamma_+({\bf t}) = \lvacn \gamma_+({\bf t}) \, J_k.
\ee
This relation allows us to represent the action of algebra $\gl(\infty)$
in terms of the differential operators.

Given a partition $\lambda$,
one can introduce the Schur polynomials (or Schur functions) 
via the
Jacobi--Trudi formulas:
\be\label{schur2}
s_{\lambda}({\bf t})=\det_{1\leq i,j\leq \ell (\lambda )}
h_{\lambda_i -i +j}({\bf t}),
\ee
where the elementary Schur polynomials $h_k({\bf t})$ are defined by
\be
e^{\sum_{k\geq 1} t_k z^k}=:\sum_{k\in \mathbb Z} h_k({\bf t}) z^k.
\ee

For the empty diagram
$s_{\emptyset}({\bf t})=1$. The action of the group elements $\gamma_\pm({\bf t})$ on the vacuum states is described by the following expansion
\begin{equation}
\begin{split}\label{lambda2}
{\gamma_-({\bf t})}\rvacn &= \sum_{\lambda} (-1)^{b(\lambda )}
s_{\lambda}({\bf t})\left |\lambda , n\rbr, \\
\lvacn {\gamma_+({\bf t})}&=\sum_{\lambda} (-1)^{b(\lambda )}
s_{\lambda}({\bf t})\lbr \lambda , n \right |,
\end{split}
\end{equation}
where 
the sums run over all partitions $\lambda$, including the empty one and $b(\emptyset )=0$.

The skew Schur functions
\be\label{lambda34}
s_{\lambda/ \mu}({\bf t})=\det_{1\leq i,j\leq \ell (\lambda )}
h_{\lambda_i -\mu_j -i+j}({\bf t})
\ee
satisfy
\be\label{Ssum}
s_\lambda({\bf t}+{\bf \tilde t})=\sum_\mu s_{\lambda/\mu}({\bf t}) s_\mu({\bf \tilde t}).
\ee
Moreover, expansions \eqref{lambda2} can be generalized as follows:
\begin{proposition}\cite[Proposition 2.11.]{AZ}\label{Propsk}
 It holds
\be\label{lambda33}
\begin{array}{l}
\displaystyle{
{\gamma_-({\bf t})}\left |\lambda , n\right > = 
\sum_{\mu} (-1)^{b(\mu/ \lambda )}
s_{\mu/ \lambda}({\bf t})\left |\mu , n\rbr},
\\ \\
\displaystyle{
{\gamma_+({\bf t})}\left |\lambda , n\right > = 
\sum_{\mu} (-1)^{b(\lambda/ \mu )}
s_{\lambda/ \mu}({\bf t})\left |\mu , n\rbr},
\\ \\
\displaystyle{
\left <\lambda , n \right |
{\gamma_+({\bf t})}=\sum_{\mu} (-1)^{b(\mu/ \lambda )}
s_{\mu/ \lambda}({\bf t})\lbr \mu , n \right |},
\\ \\
\displaystyle{
\left <\lambda , n \right |
{\gamma_-({\bf t})}=\sum_{\mu} (-1)^{b(\lambda/ \mu )}
s_{\lambda/ \mu}({\bf t})\lbr \mu , n \right |},
\end{array}
\ee
where $b(\lambda / \mu )=b(\lambda )-b(\mu )$. 
\end{proposition}

Schur functions satisfy the Cauchy--Littlewood formula
\be
e^{\sum_{k=1}^\infty k\, t_k\, \tilde t_k}=\sum_\lambda s_\lambda({\bf t}) s_\lambda({\bf \tilde t}).
\ee
Skew Schur functions satisfy a closely related summation formula
\be
e^{\sum_{k=1}^\infty k\, t_k\, \tilde t_k} \sum_\lambda s_{\mu/\lambda}({\bf \tilde t}) s_{\nu/\lambda}({\bf t})=\sum_\lambda s_{\lambda/\mu}({\bf t}) s_{\lambda/\nu}({\bf \tilde t}),
\ee
which reduces to the Cauchy--Littlewood formula for $\mu=\nu=\emptyset$.

\section{Nested hypergeometric tau-functions}\label{S3}

In this section, we introduce the nested hypergeometric tau-functions. For a general nested hypergeometric tau-function we derive 
the skew Schur function expansion and give a cut-and-join description. 

Let us consider the diagonal abelian subgroup of $\GL(\infty)$ group, which consists of the group elements of the form
\be\label{diagge}
O=\exp \left(\sum_{k \in {\mathbb  Z}} T_k \normord \psi_k \psistar_k \normord\right)
\ee
parametrized by ${\bf T}:=\{\dots,T_{-1},T_0,T_1,\dots\}$.

The fermionic Fock space states $\left |\lambda , n\right >$ and $\left <\lambda , n \right |$ are the eigenstates of these operators.
Let
\be\label{ham7}
c_n =\left \{  \begin{aligned}
&e^{-T _{-1}-T _{-2}-\ldots -T _n}, \quad &n<0,
\\
&1,  \quad \quad \quad \quad \quad \quad \quad &n=0,
\\
&e^{T _{0}+T _{1}+\ldots + T_{n-1}}, \quad \quad \!  &n>0,
 \end{aligned}
\right.
\ee
be the solution to the difference equation $c_{n+1}/c_{n}=e^{T_n}$ fixed by $c_0=1$. Consider the {\em content product}
\be
r_{\lambda,n}=\prod_{(i,j)\in \lambda} G(n+j-i),
\ee
where we take the product over all cells of the Young diagrams associated to a partition $\lambda$ with the factors dependent on the {\em content} $j-i$. Here the {\em weight generating function} $G(z)$ is related to the parameters $T_j$ as follows:
\be
G(i):=e^{T_i-T_{i-1}}.
\ee
Equivalently
\be\label{ham8}
r_{\lambda,n}=e^{\sum_{i=1}^{d(\lambda )}T _{n+\alpha_i} 
- T_{n-\beta_i-1}}= e^{\sum_{j\geq 1}T_{n+\lambda_j -j} - T_{n-j}}.
\ee

Then from the definition of the states $\left |\lambda , n\right >$ and $\left <\lambda , n \right |$ and \eqref{ham8} we immediately have the following statement. 
\begin{lemma}\label{prop3.1}
For a diagonal group element $O$ we have
\begin{equation}
\begin{split}
\label{Oact}
O\left |\lambda , n\right > &=c_n\, r_{\lambda,n}  \left |\lambda , n\right >, \\
\left <\lambda , n \right | O &=c_n\,  r_{\lambda,n} \left <\lambda , n \right |. 
\end{split}
\end{equation}
\end{lemma}

Let $m \geq 0$ be a non-negative integer. Let us choose an ordered sequence of signs, $\sigma=(\sigma_m,\sigma_{m-1},\dots,\sigma_1)$, where $\sigma_i$ is a sign, that is $\sigma_i=+$ or $\sigma_i=-$. There are exactly $2^{m}$ different sequences, for example, for $m=2$ we have 4 possibilities: $\sigma=++$, $\sigma=+-$, $\sigma=-+$, or $\sigma=--$.

With $0 \leq k \leq m+1$ we associate infinite sets of the formal variables, ${\bf t_{k}}:=\{t_{1,k},t_{2,k},\dots\}$. We also introduce the parameters ${\bf T^{(k)}}:=\{\dots,T_{-1}^{(k)},T_{0}^{(k)},T_{1}^{(k)},\dots\}$  for $0\leq k \leq m$ and the diagonal group elements
\be
O_k=\exp \left(\sum_{j \in {\mathbb  Z}} T_j^{(k)} \normord \psi_j \psistar_j \normord\right).
\ee
Let us stress that all variables ${\bf t_k}$ and ${\bf T}^{(k)}$ are independent.

\begin{definition}\label{def1}
For any $\sigma$ and ${\bf T^{(k)}}$ the nested hypergeometric tau-function is given by the vacuum expectation value
\be\label{taudef}
\tau_n^{m,\sigma}({\bf t_{m+1}},\dots,{\bf t_0})
=\lvacn \gamma_+({\bf t_{m+1}})\, O_{m} \,\gamma_{\sigma_m}({\bf t_{m}}) \,O_{m-1} \dots  O_{1}\, \gamma_{\sigma_1}({\bf  t_{1}})\,
O_{0}\,  \gamma_-({\bf  t_{0}}) \rvacn.
\ee
\end{definition}
This is a formal series in the variables ${\bf t_k}$. Since $O$ and $\gamma_\pm({\bf t})$ belong to $\GL(\infty)$ group,
this is a 2D Toda lattice tau-function in the variables ${\bf t_0}$, ${\bf t_{m+1}}$, and $n$, while ${\bf t_{1}},\dots,{\bf t_m}$, ${\bf T^{(0)}},\dots,{\bf T^{(m)}}$, and $\sigma$ are the parameters of the tau-function.

\subsection{Skew Schur function expansion}

Let us find the expansion of the tau-function $\tau_n^{m,\sigma}$. We use Proposition \ref{Propsk} and Lemma \ref{prop3.1}. 

We consider the vacuum expectation value in Definition \ref{def1} from left to right. The initial state is $ \gamma_-({\bf  t_{0}}) \rvacn= \sum_{\lambda} (-1)^{b(\lambda)}
s_{\lambda}({\bf t_{0}})
\left |\lambda , n \right > $. Then, according to Proposition \ref{Propsk} insertion of the operator ${\gamma_\pm({\bf t})}$ creates an additional sum over partitions with the skew Schur function.  By Lemma \ref{prop3.1} insertion of the diagonal operator $O$ creates a partition dependent weight $r_{\lambda,n}$.

Let us consider the simplest example, which corresponds to the case $m=0$,
\be\label{HHtf}
 \tau_n^{0,\emptyset}=\lvacn {\gamma_{+}({\bf  t_{1}})} \,
O_{0} \,  {\gamma_-({\bf  t_{0}})}\rvacn= c_n^{(0)}\sum_\lambda r^{(0)}_{\lambda,n} s_\lambda({\bf t_{1}}) s_\lambda({\bf t_{0}}).
\ee
These are hypergeometric tau-functions \cite{KMMM,OS}, which describe weighted Hurwitz numbers \cite{O,P,GJ2,GPH}.

For $m=1$ we have two possible values of $\sigma$. Corresponding tau-functions are
\begin{equation}
\begin{split}\label{1+}
 \tau_n^{1,+}&=\lvacn {\gamma_{+}({\bf  t_{2}})} \, O_{1} \, {\gamma_{+}({\bf  t_{1}})} \, O_{0} \,  {\gamma_-({\bf  t_{0}})}\rvacn\\
 &= c_n^{(0)}c_n^{(1)}\sum_{\lambda_0,\lambda_1}s_{\lambda_1}({\bf t_2})\, r^{(1)}_{\lambda_1,n}\, s_{\lambda_0/\lambda_1}({\bf t_1})\, r^{(0)}_{\lambda_0,n}\,  s_{\lambda_0}({\bf t_0})
\end{split}
\end{equation}
and
\begin{equation}
\begin{split}\label{1-}
 \tau_n^{1,-}&=\lvacn {\gamma_{+}({\bf  t_{2}})} \, O_{1} \, {\gamma_{-}({\bf  t_{1}})} \, O_{0} \, {\gamma_-({\bf  t_{0}})}\rvacn\\
& =c_n^{(0)}c_n^{(1)}\sum_{\lambda_0,\lambda_1}s_{\lambda_1}({\bf t_2})\, r^{(1)}_{\lambda_1,n}\, s_{\lambda_1/\lambda_0}({\bf t_1})\, r^{(0)}_{\lambda_0,n}\,  s_{\lambda_0}({\bf t_0}).
\end{split}
\end{equation}
These tau-functions and their relation to other models are discussed in Section \ref{S5.2}.

It is easy to see the general structure of the expansion. 
Let us introduce the notation $(\lambda/\mu)^+:=\lambda/\mu$ and $(\lambda/\mu)^-:=\mu/\lambda$. The signs $(-1)^{b(\lambda/\mu)}$ from \eqref{lambda33} are always compensated in the final expression and never contribute. Then for the nested hypergeometric tau-function $\tau_n^{m,\sigma}$ we have the following skew Schur function expansion.
\begin{theorem}\label{T1}
\be\label{MTS}
\tau_n^{m,\sigma}({\bf t_{m+1}},\dots,{\bf t_0})=\prod_{j=0}^m c_n^{(j)}\sum_{\lambda_0,\dots,\lambda_m} s_{\lambda_m}({\bf t_{m+1}})  \prod_{i=1}^m \left(r^{(i)}_{\lambda_i,n}  s_{(\lambda_{i-1}/\lambda_i)^{\sigma_i}}({\bf t_{i}})\right) \, r^{(0)}_{\lambda_0,n} \, s_{\lambda_0}({\bf t_0}).
\ee
\end{theorem}
This family generalizes the hypergeometric tau-functions \eqref{HHtf} in a natural way. While in the expansion of the hypergeometric tau-function \eqref{HHtf} the summation runs over all partitions, in the expansion \eqref{MTS} only some combinations of partitions survive.

We say that $\lambda\subset \mu$ if $\lambda_i\leq \mu_i$ for all $i$. 
The non-trivial terms in the sum in \eqref{MTS} are given by sequences of nested partitions. For instance, if $\sigma=(+,+,+\dots)$, then only the nested sequences of partitions $\lambda_m\subset \lambda_{m-1} \subset \dots \subset \lambda_{0}$ contribute to the summation. For other choices of the signs $\sigma$ the nested structure is more involved, namely $\lambda_i \subset \lambda_{i-1}$ if $\sigma_i=+$ and $\lambda_i \supset \lambda_{i-1}$ if $\sigma_i=-$. Because of this nested structure we call $\tau_n^{m,\sigma}({\bf t_{m+1}},\dots,{\bf t_0})$ the nested hypergeometric tau-function. 

Below, depending on the context, we consider the nested hypergeometric tau-functions
 of the 2D Toda lattice, 2-component KP (or MKP) hierarchy or even 1-component KP (or MKP) hierarchy. Therefore, by the the nested hypergeometric tau-functions we always mean \eqref{taudef} as a function of a proper set of variables and parameters. 


Let $\sigma^*=(-\sigma_1,-\sigma_{2},\dots,-\sigma_{m})$. Then 
\be\label{ptom}
\tau_n^{m,\sigma}({\bf t_{m+1}},\dots,{\bf t_0})=\tau_n^{m,\sigma^*}({\bf t_0},\dots,{\bf t_{m+1}}),
\ee 
where for the right hand side tau-function the order of all group operators $O$ and $\gamma_\pm$ and the signs of operators $\gamma_\pm$ is inverted:
\be
\tau_n^{m,\sigma^*}({\bf t_0},\dots,{\bf t_{m+1}})=\lvacn \gamma_+({\bf  t_{0}}) \, O_{0} \, \gamma_{-\sigma_1}({\bf  t_{1}})\,  O_{1}\, \dots  O_{m-1} \gamma_{-\sigma_m}({\bf t_{m}}) \, O_{m} \,
\gamma_-({\bf t_{m+1}}) \rvacn.
\ee
In particular, 
\eqref{1+} and \eqref{1-} describe the same families of tau-functions and it is enough to consider only one of them. If we identify the nested hypergeometric tau-functions, related by this duality, then there are $2^{m-1}$ inequivalent possibilities for odd $m$ and $2^{m-1}+2^{m/2-1}$ inequivalent possibilities for even $m$.

The nested structure of the considered tau-functions leads to the simple reduction properties. Namely, the nested hypergeometric tau-function $\tau_n^{m,\sigma}$ reduces to a tau-function of the form $\tau_n^{m-1,\sigma'}$ if ${\bf t_{j}}={\bf 0}$,
or if $O_j=1$ and $\sigma_j=\sigma_{j+1}$  for some $j$  (we assume that $\sigma_0=-$ and $\sigma_{m+1}=+$). In this last case $\tau_n^{m-1,\sigma'}$ depends only on the combination ${\bf t_j+t_{j+1}}$.

\subsection{Cut-and-join operators}\label{S3.2}
Let us describe the nested hypergeometric tau-functions $\tau_n^{m,\sigma}$ by the cut-and-join operators. As a result, the nested hypergeometric tau-functions are constructed recursively in $m$.
There are several types of the cut-and-join descriptions, and another option convenient for certain families of the diagonal group operators is discussed in Section \ref{S4.4}.

Recall Lemma \ref{prop3.1}. For a diagonal group element $O$ consider the differential operators $\widehat{O}$ with the Schur functions as the eigenfunctions, 
\be
\widehat{O}({\bf t},n)\cdot s_\lambda({\bf t}) =c_n r_{\lambda,n}  s_\lambda({\bf t}),
\ee
then
\be
\widehat{O}({\bf t},n)\cdot \lvacn \gamma_+({\bf t}) = \lvacn \gamma_+({\bf t}) \, O
\ee
and
\be\label{WE}
\widehat{O}({\bf t},n)\cdot \gamma_-({\bf t}) \rvacn =  O \, \gamma_-({\bf t})  \rvacn.
\ee

These operators can be constructed from the boson-fermion correspondence in terms of the vertex operators. If $T_k =p(k)$ for some polynomial $p(z)$, then the operator $\widehat{O}$ has  a nice description in terms of a finite combination of the cut-and-join operators operators $\widehat{C}_k$,  $\widehat{O}=\exp \left(\sum \alpha_k \widehat{C}_k\right)$ \cite{AMMN}.

The cut-and-join description of a nested hypergeometric tau-functions $\tau_n^{m,\sigma}$ can be constructed recursively. Namely, we have
\begin{equation}
\begin{split}
\tau_n^{m,\sigma}&=\lvacn \gamma_+({\bf t_{m+1}})\, O_{m} \, \gamma_{\sigma_m}({\bf t_{m}}) \, O_{m-1} \dots  O_{1}\, \gamma_{\sigma_1}({\bf  t_{1}})\,
O_{0} \, \gamma_-({\bf  t_{0}}) \rvacn\\
&= \widehat{O}_{m}({\bf t_{m+1}},n)\cdot \lvacn \gamma_+({\bf t_{m+1}})\,  \gamma_{\sigma_m}({\bf t_{m}}) O_{m-1} \dots  O_{1}\, \gamma_{\sigma_1}({\bf  t_{1}})
O_{0} \, \gamma_-({\bf  t_{0}}) \rvacn.
\end{split}
\end{equation}
If $\sigma_m=-$, we have
\begin{equation}
\begin{split}
 \lvacn \gamma_+({\bf t_{m+1}}) \gamma_{-}({\bf t_{m}}) 
=e^{\sum_{k=1}^\infty k\, t_{m+1,k}\, t_{m,k}}\, \lvacn \gamma_+({\bf t_{m+1}}),
\end{split}
\end{equation}
therefore
\be\label{m-1+}
\tau_n^{m,\sigma}({\bf t_{m+1}},\dots,{\bf t_0})=\widehat{O}_{m}({\bf t_{m+1}},n)\cdot  e^{\sum_{k=1}^\infty k \,t_{m+1,k}\, t_{m,k}}\, \tau_n^{m-1,\sigma'}({\bf t_{m+1}},{\bf t_{m-1}},\dots,{\bf t_0}),
\ee
where $\sigma'=(\sigma_{m-1},\dots,  \sigma_1)$ and
\be\label{mina}
 \tau_n^{m-1,\sigma'}({\bf t_{m+1}},{\bf t_{m-1}},\dots,{\bf t_0})=\lvacn \gamma_+({\bf t_{m+1}})\, O_{m-1} \, \gamma_{\sigma_{m-1}}({\bf t_{m-1}})  \dots  O_{1}\, \gamma_{\sigma_1}({\bf  t_{1}})
O_{0} \, \gamma_-({\bf  t_{0}}) \rvacn.
\ee

If $\sigma_m=+$, we have
\begin{equation}
\begin{split}
 \lvacn \gamma_+({\bf t_{m+1}}) \gamma_{+}({\bf t_{m}})=  \lvacn \gamma_+({\bf t_{m+1}+t_{m}}),
\end{split}
\end{equation}
and
\be\label{m-1-}
\tau_n^{m,\sigma}({\bf t_{m+1}},\dots,{\bf t_0})=\widehat{O}_{m}({\bf t_{m+1}},n) \cdot \tau_n^{m-1,\sigma'}({\bf t_{m+1}+t_m},{\bf t_{m-1}},\dots,{\bf t_0})
\ee
and $\tau_n^{m-1,\sigma'}$ is given by \eqref{mina}.

Equations \eqref{m-1+} and \eqref{m-1-} describe a recursion in $m$, and allow us to reduce the description of the tau-functions $\tau_n^{m,\sigma}$ to the action of the group operators $\widehat{O}$ on the elementary exponential functions. 
There are equivalent descriptions if we start from the right vacuum state $\rvacn$ and a group element $O_{0}$ with \eqref{WE} or combine two directions, see for instance \eqref{o1}--\eqref{o3} for all possibilities for $m=1$.

\section{Matrix models}\label{S4}
In this section, we construct matrix models for all nested hypergeometric tau-functions with the diagonal group elements described by the weight generating functions \eqref{Gcovered}. We also derive a cut-and-join description of the nested hypergeometric tau-functions with certain combinations of diagonal group operators associated with rational weight generating functions. 

\subsection{Elementary building blocks}

Let us recall  the matrix models which serve as elementary building blocks for the construction of the nested hypergeometric tau-functions, introduced in the previous section.
The Schur function expansion of these basic matrix models is well known \cite{ZJZ,Orlov,O1,MMRP,AMMN}, and the main goal of this section is to provide a brief reminder. 

We assume that the matrix integrals are over the space of the $N\times N$ matrices if it is not specified explicitly.
We work only with formal matrix models, and, by definition, consider the partition functions of the matrix models by the perturbative expansion of the matrix integrals in the corresponding variables.

We focus on the matrix integrals, that produce the following weight generating functions:
\be\label{G}
G^+(z)=1+uz,\quad \quad G^-(z)=\frac{1}{1+uz}, \quad \quad G^{\exp}(z)=e^{uz}.
\ee
These functions are associated with the strictly monotone Hurwitz numbers, monotone Hurwitz numbers, and simple Hurwitz numbers correspondingly, see, e.g., \cite{Ferm}. 
The content products
\be
r^\bullet_\lambda(u)=\prod_{(i,j)\in \lambda}G^\bullet(j-i)
\ee
for $\bullet = +, -,  \exp$ are the corresponding weights of the Schur expansions of the hypergeometric tau-functions \eqref{HHtf} with $n=0$,
and  we denote by $O_\bullet(u)$ the  diagonal group elements associated with them. For $+$ and $-$ indices we also have a reciprocity relation
\be
r^-_\lambda(u)=\frac{1}{r^+_\lambda(u)},
\ee
and, therefore, $O_+(u)=O^{-1}_-(u)$. For $N\in {\mathbb Z}_{>0}$ the coefficients $r^\pm_\lambda(N^{-1})$ can be represented as a ratio of the Schur functions at a particular locus of the arguments,
\be\label{Rassch}
r^+_\lambda(N^{-1})=N^{-|\lambda|}\frac{s_\lambda(t_k={N}/{k})}{s_\lambda(t_k=\delta_{k,1})}.
\ee

\begin{remark}
Any weight generating function given by a formal series, $G(z)\in 1+{\mathbb C}[\![z]\!]$, can be described by certain normal matrix integrals, see \cite{O2} and  \cite{MMRP} for two different constructions. In this paper we restrict ourself to the finite products of the weight generating functions \eqref{G} because they can be expressed in terms of the Gaussian integrals and elementary residues, and, therefore, can be investigated and computed explicitly.
\end{remark}

The sum with the weight generating function $G^+$ can be described by the complex matrix integral (see, e.g., (77) in \cite{Orlov})
\be\label{sumpl}
\sum_{\ell(\lambda)\leq N} s_\lambda({\bf t})s_\lambda({\bf \bar t}) r_\lambda^{+}(N^{-1})=\int_\mathfrak{C} \left[d \mu_\mathfrak{C}({Z})\right]  \exp\left(\sum_{k=1}^\infty t_k\Tr{{Z}}^k+\bar{t}_k\Tr{ {Z^\dagger }}^k\right).
\ee
Here we integrate over the space of the $N\times N $ complex matrices with the measure 
\be
\left[d\mu_\mathfrak{C} ({Z})\right] =  \left[d {Z}\right]  \exp\left(-N\,  \Tr { ZZ^\dagger}\right),
\ee
where $ \left[d {Z}\right]=c_\mathfrak{C} \prod_{i,j=1}^N d Z_{ij} d\bar{Z}_{ij}$ is the flat measure on the space of complex matrices
 with the constant $c_\mathfrak{C}$ specified by the normalization condition 
$\int_\mathfrak{C} \left[d\mu_\mathfrak{C} ({Z})\right]=1$. 

The coefficients of the series expansion in ${\bf t}$ and ${\bf \bar t}$ variables of the right hand side of \eqref{sumpl} are given by the Gaussian integrals of the polynomial functions and can be evaluated explicitly. 

The weight generating function $G^-$ can be described by the double unitary integral. To describe it we first remind the well known orthogonality property of the Schur functions, which can be found in \cite[Section VII]{Macdonald}, and in a more general form that we need, in \cite[Lemma 2.1]{Novak}. 
\begin{lemma}\label{LemmaU}
For any partitions $\lambda,\mu$ and matrices $A,B \in \Mat_N({\mathbb C})$, we have
\be\label{orth0}
\int_\mathfrak{U}\left[d{U}\right] s_\lambda(AUBU^\dagger)=\frac{s_\lambda(A)s_\lambda(B)}{s_\lambda(t_k={N}/{k})}
\ee
and
\be\label{orth}
\int_\mathfrak{U}\left[d{U}\right] s_\lambda(AU)s_\mu(BU^\dagger)=\delta_{\lambda \mu}\frac{s_\lambda(AB)}{s_\lambda(t_k={N}/{k})}.
\ee
\end{lemma}
Here the Haar measure $\left[d{U}\right]$ on the unitary group $U(N)$ is normalized by $\int_\mathfrak{U} \left[d{U}\right]=1$. Then, form \eqref{orth} and the Cauchy--Littlewood formula it immediately follows that
\be\label{Uele}
\sum_{\ell(\lambda)\leq N} s_\lambda({\bf t})s_\lambda({\bf \bar t})  r_\lambda^{-}(N^{-1})= \int_{\mathfrak{U}^2} \left[d\mu_\mathfrak{U}{(U)}\right] \exp\left(\sum_{k=1}^\infty t_k \Tr {U^\dagger}^k+ {\bar t}_k \Tr \tilde{U}^k \right).
\ee
Here we integrate over two independent unitary matrices $U$ and $\tilde U$ and
\be
\left[d\mu_\mathfrak{U}{(U)}\right]=  \left[d{U}\right]   [d{\tilde U}] \exp( N \, \Tr U \tilde U^\dagger).
\ee
Applying the Harish-Chandra--Itzykson--Zuber formula one can reduce this matrix integral to the eigenvalue one. 
Coefficients of the ${\bf t}$ and ${\bf \bar t}$ series expansion of this eigenvalue integral are given by combinations of the elementary residues. 

Let us recall that a matrix $Z$ is normal if it commutes with its Hermitian conjugate, $[Z,Z^\dagger]=0$.
For the case $G^{\exp}$ there is a normal matrix integral description (see (11) in \cite{MMRP})
\be\label{log}
\sum_{\ell(\lambda)\leq N} s_\lambda({\bf t})s_\lambda({\bf \bar t}) r_\lambda^{\exp}(u)=\int_\mathfrak{N} \left[d \mu_{\mathfrak{N}}({Z})\right]
\exp\left(\sum_{k=1}^\infty t_k \Tr {Z}^k+ \bar t_k \Tr {Z}^{\dagger k}\right).
\ee
Here the integral is taken over the space of $N\times N$ normal matrices with complex eigenvalues.
 A normal matrix $Z$ can be diagonalized as $Z=UzU^\dagger$,
with unitary $U$ and diagonal $z=\diag(z_1,\dots,z_N)$ with complex eigenvalues, then the flat measure on the space of normal matrices is given by $\left[d {Z} \right]= [dU] |\Delta(z)|^2 \prod_{i=1}^N dz_i \, d\bar{z}_i$. Here $\Delta(z)$ is the Vandermonde determinant,
\be
\Delta(z)=\prod_{i<j}(z_j-z_i).
\ee
The measure in \eqref{log} is given by
\be
\left[d \mu_{\mathfrak{N}}({Z})\right]=C_{\mathfrak{N}} \frac{\left[d {Z} \right]}{\left(\det {Z Z^\dagger}\right)^{N+\frac{1}{2}} }
\exp\left(-\frac{1}{2u}\Tr \log^2 {Z Z^\dagger}\right),
\ee
where $C_{\mathfrak{N}}$, dependent on $N$ and $u$, is fixed by the normalization constraint $\int_\mathfrak{N} \left[d \mu_{\mathfrak{N}}({Z})\right]=1$. Integral \eqref{log} can be immediately reduced to the eigenvalue integral which, after a change of integration variables, is a combination of the Gaussian ones \cite{MMRP}.

Let us stress that the above matrix integral descriptions are not unique, for example the right hand side of  \eqref{sumpl} has an alternative description by a two-matrix model
\be\label{double}
\sum_{\ell(\lambda)\leq N} s_\lambda({\bf t})s_\lambda({\bf \bar t}) r_\lambda^{+}(N^{-1})=\int_{\mathfrak{H}^2} \left[d{X}\right]  \left[d{Y}\right]\exp\left(\sqrt{-1} N \Tr {XY}+\sum_{k=1}^\infty t_k\Tr{X^k}+\bar{t}_k\Tr{Y}^k\right),
\ee
where the integration is over the space of a pair of Hermitian matrices $X$ and $Y$ with the flat measure $ \left[d{X}\right]={c}_{\mathfrak{H}} \prod_{i<j} d\Im \Phi_{ij} \,d\Re \Phi_{ij}\prod_{i=1}^n d \Phi_{ii}$ and a constant ${c}_{\mathfrak{H}}$ fixed by a normalization constraint
$\int_{\mathfrak{H}^2} \left[d{X}\right]  \left[d{Y}\right]\exp\left(-N \sqrt{-1} \Tr {XY}\right)=1$. 
A Hermitian matrix $X$ can be diagonalized, $X=UxU^\dagger$, with a unitary $U$ and real diagonal $x=\diag(x_1,\dots,x_N)$, then the flat measure on the 
space of Hermitian matrices is $\left[ dX\right]=\tilde{c}_{\mathfrak{H}}\left[ dU\right] \Delta(x)^2 \prod_{i=1}^N d x_i$ for another constant $\tilde{c}_{\mathfrak{H}}$.

For a function $f({\bf t})$ and a matrix $A$ let us define the {\em Miwa parametrization} $f(A)=f({\bf t})\big|_{t_k=\frac{1}{k}\Tr A^k}$.
Then, in the Miwa parametrization the sum given by \eqref{Uele} can be represented as a unitary matrix integral
\be\label{unit2}
\sum_{\ell(\lambda)\leq N} s_\lambda(A)s_\lambda({\bf t}) r_\lambda^{-}(N^{-1})= \int_\mathfrak{U} \left[d{U}\right] \exp\left(N \Tr{U^\dagger A}+\sum_{k=1}^\infty t_k \Tr U^k  \right)
\ee
or the Harish-Chandra--Itzykson--Zuber matrix integral 
\be\label{unit}
\sum_{\ell(\lambda)\leq N} s_\lambda(A)s_\lambda(B) r_\lambda^{-}(N^{-1})= \int_\mathfrak{U} \left[d{U}\right] \exp\left(N \Tr({ {UAU^\dagger B}}) \right).
\ee
Both identities follow form Lemma \ref{LemmaU} and the Cauchy--Littlewood formula.

\begin{remark}
These examples indicate an important  general property of the matrix models we consider in this paper -- usually there are several equivalent matrix models for a given tau-function. In the next section we provide a universal recipe for the matrix integral description of the nested hypergeometric tau-functions $\tau_0^{m,\sigma}$, however, sometimes the ambiguity can help us to simplify the matrix integral expressions. In particular, the Miwa parametrization, natural for the matrix models, often leads to simplifications. We will return to this ambiguity in Section \ref{S5}.
\end{remark}

\begin{remark}
In the above formulas we use a normalization natural for the applications related to the Hurwitz numbers.
For other applications it could be more convenient to consider other normalizations of the time variables, consistent with the 2D Toda hierarchy, in particular,
\begin{multline}
\sum_{\ell(\lambda)\leq N} s_\lambda({\bf t})s_\lambda({\bf \bar t}) \prod_{(i,j)\in \lambda}(N-i+j)\\
=N^{-{N^2}}\int_{\mathfrak{H}^2} \left[d{X}\right]  \left[d{Y}\right]\exp\left(\sqrt{-1}\Tr {XY}+\sum_{k=1}^\infty t_k\Tr{X^k}+\bar{t}_k\Tr{Y}^k\right)
\end{multline}
and
\be
\sum_{\ell(\lambda)\leq N} \frac{s_\lambda(A)s_\lambda(B) }{ \prod_{(i,j)\in \lambda}(N-i+j)}= \int_\mathfrak{U} \left[d{ U}\right] \exp\left(\Tr( {UAU^\dagger B}) \right).
\ee
These  formulas are related to \eqref{double} and \eqref{unit} by a simple rescaling of the ${\bf t}$ or $A$ variables. 
\end{remark}

\subsection{Chains of matrix integrals}\label{S4.2}

Any nested hypergeometric tau-function $\tau_n^{m,\sigma}$, given by Definition \ref{def1}, if all its diagonal group elements $O_j$ correspond to the weight generating functions of the form
\be\label{Gcovered}
G(z)=\frac{\prod_{i=1}^\ell(1+u_iz)}{\prod_{j=1}^n(1+v_jz)}e^{wz}
\ee
for some finite $\ell$ and $n$,
can be described by a multi-matrix integral with the chain structure. We restrict ourselves to the zero charge sector $n=0$; other cases can be described similarly. 
\begin{remark}
It is also possible to introduce in $G$ an arbitrary constant factor $Q$, which corresponds to the weight $Q^{|\lambda|}$,
by a rescaling of the ${\bf t}$ variables. We will not consider this rescaling below. 
\end{remark}
 
We start with the fermionic vacuum expectation value \eqref{taudef} and consider it from right to  left. With a state $\left |\lambda\rbr=\left |\lambda, 0\rbr$ we associate a Schur function $s_\lambda({\bf \tilde t})$ with auxiliary variables ${ \tilde t_k}$, which can be identified with $J_{-k}/k$. Then the function $\exp\left( \sum_{k=1}^\infty k \tilde{t}_k t_{k}  \right)=\sum_{\lambda} s_\lambda({\bf t}) s_\lambda({\bf \tilde t})$ is associated with the state ${\gamma_-({\bf  t})} \rvac$. 

To construct matrix models we need to describe the action of the group elements $O_\pm(u)$, $O_{\exp}(u)$, and  $\gamma_\pm({\bf t})$. For $\ell(\lambda)\leq N$ insertion of weights $r_\lambda^{\pm}$ and $r_\lambda^{\exp}$ corresponds to the matrix integrals 
\begin{equation}\label{mmap}
\begin{split}
\exp\left(\sum_{k=1}^\infty k \tilde t_k t_k\right) \,\,\, &\mapsto \,\,\, \sum_{\ell(\lambda)\leq N} s_\lambda({\bf t})s_\lambda({\bf \tilde t}) r_\lambda^{+}(N^{-1})=\int_\mathfrak{C} \left[d \mu_\mathfrak{C}({Z})\right]  \exp\left(\sum_{k=1}^\infty t_k\Tr{{Z}}^k+\tilde{t}_k\Tr{ {Z^\dagger }}^k\right),\\
\exp\left(\sum_{k=1}^\infty k \tilde t_k t_k\right) \,\,\, &\mapsto \,\,\, \sum_{\ell(\lambda)\leq N} s_\lambda({\bf t})s_\lambda({\bf \tilde t})  r_\lambda^{-}(N^{-1})= \int_{\mathfrak{U}^2} \left[d\mu_\mathfrak{U}{(U)}\right] \exp\left(\sum_{k=1}^\infty t_k \Tr {U^\dagger}^k+ {\tilde t}_k \Tr \tilde{U}^k \right),\\
\exp\left(\sum_{k=1}^\infty k \tilde t_k t_k\right) \,\,\, &\mapsto \,\,\,  \sum_{\ell(\lambda)\leq N} s_\lambda({\bf t})s_\lambda({\bf \tilde t}) r_\lambda^{\exp}(u)=\int_\mathfrak{N} \left[d \mu_{\mathfrak{N}}({Z})\right]\exp\left(\sum_{k=1}^\infty t_k \Tr {Z}^k+ \tilde t_k \Tr {Z}^{\dagger k}\right).
 \end{split}
\end{equation}
These transformations preserve the exponential form of the dependence of the ${\bf \tilde t}$ variables.

As a result of application of any combination of these matrix integrals we have the dependence on ${\bf \tilde t}$ variables in the form $\exp\left(\sum_{k=1}^\infty \tilde t_k \Tr {\Phi}^k\right)=\sum_\lambda s_\lambda({\bf \tilde t})s_\lambda(\Phi)$ for some matrix $\Phi$. 
According to \eqref{Ssum}, insertion of the operators $\gamma_+({\bf  \bar t})$ and $\gamma_-({\bf  \bar t})$ in terms of the auxiliary variables ${\bf \tilde t}$
yields correspondingly
\begin{equation}\label{gammap}
\begin{split}
\exp\left(\sum_{k=1}^\infty k \tilde t_k t_k\right) \quad &\mapsto \quad  \sum_\mu s_{\lambda/\mu}({\bf  \bar t}) s_\mu({\bf \tilde t}) s_\lambda({\bf t}) = \exp\left(\sum_{k=1}^\infty k(\tilde t_k+ \bar{t}_{k}) t_k\right),\\
\exp\left(\sum_{k=1}^\infty k \tilde t_k t_k\right) \quad &\mapsto \quad \sum_{\mu,\lambda} s_{\mu/\lambda}({\bf  \bar t}) s_\mu({\bf \tilde t})s_\lambda({\bf t})=\exp\left(\sum_{k=1}^\infty k \tilde t_k ( t_k+ \bar{t}_{k})\right).
\end{split}
\end{equation}
Both preserve the exponential form of the dependence on the auxiliary variables ${\bf \tilde t}$.

At the final step we identify ${\bf \tilde t}$ with ${\bf t_{m+1}}$. As a result, we obtain a multi-matrix integral construction which describes all tau-functions $\tau_0^{m,\sigma}$ with any combination of the weight generating functions \eqref{Gcovered}.
\begin{proposition}\label{prop4.1}
All nested hypergeometric tau-functions \eqref{taudef} with diagonal group operators $O_j$ associated with the weight generating functions \eqref{Gcovered} are described by the multi-matrix models, which are combinations
of the complex, unitary, and normal matrix integrals \eqref{sumpl}, \eqref{Uele}, and \eqref{log}, namely, with each diagonal operator $O_j$ corresponding to the weight function \eqref{Gcovered} we associate $\ell$ integrals \eqref{sumpl}, $n$ integrals \eqref{Uele}, and, if $w\neq 0$, one integral \eqref{log}.
Moreover, these multi-matrix models have a chain structure with a pairwise interaction, and can be reduced to the eigenvalue integrals. 
\end{proposition}

Vice versa, any matrix integral constructed by the subsequent application of  \eqref{mmap} and \eqref{gammap} describes a nested hypergeometric tau-function. This proposition will be illustrated with examples in Section \ref{S5}.

The dependence on all ${\bf t_j}$ variables in the constructed matrix models is explicit of the form $\exp \left(\sum_k t_{j,k}\dots\right)$, where by $\dots$ we denote linear combinations of other variables $k\,  t_{i,k}$ for $i \neq j$ and traces of the integration matrices $\Tr \Phi^k$.

Two different forms of the same nested hypergeometric tau-functions, given by our construction for two sides of \eqref{ptom}, provide essentially the same chain of the matrix integrals. In Section \ref{S5} below we provide examples which describe construction of matrix models for the nested hypergeometric tau-functions and possible ambiguity.

\begin{remark}
The matrix model construction of this section is closely related to the approach developed by Orlov \cite{Orlov,O1,Orlovchain}. 
\end{remark}

\begin{remark}
Considered multi-matrix models for the nested hypergeometric tau-functions have a chain structure. It is possible to consider more general matrix integrals associated with other graphs.  
Namely, a trivalent vertex (a pair of pants) can be described by a unitary matrix integral
\be
\int_\mathfrak{U} \left[d{U}\right] \exp\left({\sum_{k=1}^\infty t_k\Tr({ {UAU^\dagger B}})^k}\right)=\sum_{\ell(\lambda)\leq N}\frac{s_\lambda({\bf t})s_\lambda({ A})s_\lambda({B})}{s_\lambda(t_k=N/k)},
\label{vertun}
\ee
or a complex matrix integral
\be
\int_\mathfrak{C} \left[d{Z}\right]  \exp\left(-\Tr  {ZZ^\dagger}+\sum_{k=1}^\infty t_k\Tr({{ZAZ^\dagger B}})^k\right)=\sum_{\ell(\lambda)\leq N}\frac{s_\lambda({\bf t})s_\lambda({ A})s_\lambda({B})}{s_\lambda(t_k=\delta_{k,1})},
\label{comps}
\ee
see \cite{MMRP} for more details.  In general, for such generalizations the matrix models are not immediately reducible to the eigenvalue integrals anymore, and the KP/Toda integrability is broken (see also Section \ref{S6}).
\end{remark}

\subsection{Superintegrability}

\label{rmk4.2} 
Equations \eqref{sumpl}, \eqref{Uele}, and \eqref{log} can be interpreted as examples of the of matrix model superintegrability \cite{Susa,WLZZ}, namely for $\ell(\lambda)\leq N$ the integrals of the Schur functions are proportional to the same Schur functions:
\begin{equation}\label{elesu}
\begin{split}
 s_\lambda({\bf t})\, r_\lambda^{+}(N^{-1})&=\int_\mathfrak{C} \left[d\mu_\mathfrak{C} ({Z})\right]\, s_\lambda(Z^\dagger) \exp\left(\sum_{k=1}^\infty  t_k \Tr{Z}^k\right),\\
 s_\lambda({\bf  t})\, r_\lambda^{-}(N^{-1})&= \int_\mathfrak{U^2} \left[d\mu_\mathfrak{U}{(U)}\right] \, s_\lambda(U^\dagger)  \exp\left(\sum_{k=1}^\infty t_k \Tr{\tilde U}^k\right),\\
s_\lambda({\bf  t})\, r_\lambda^{\exp}(u)&=\int_\mathfrak{N} \left[d \mu_{\mathfrak{N}}({Z})\right]\, s_\lambda({Z}^{\dagger}) \exp\left(\sum_{k=1}^\infty t_k \Tr {Z}^k\right),
 \end{split}
\end{equation}
and the same equations hold true if we permute the roles of $Z$ and $Z^\dagger$, and $U$ and $U^\dagger$.

As a corollary of Proposition \ref{prop4.1} we get a general superintegrability rule for the chain of the matrix models associated with the nested hypergeometric tau-functions \eqref{taudef} with the weight generating functions of the form \eqref{Gcovered}. For simplicity, let us assume that the diagonal group element $O_m$ is non-trivial. Then the general construction of the previous section leads to a chain of matrix integrals, and the last matrix of integration, which couples to the variables ${\bf t_{m+1}}$,  we denote by $\Phi$ (that is, $\Phi$ is unitary, complex, or normal). Then
\be
\left<s_{\lambda_m}(\Phi)\right>= r^{(m)}_{\lambda_m}
\sum_{\lambda_0,\dots,\lambda_{m-1}}  \prod_{i=1}^m r^{(i-1)}_{\lambda_{i-1}}  s_{(\lambda_{i-1}/\lambda_i)^{\sigma_i}}({\bf t_{i}})\, s_{\lambda_0}({\bf t_0}),
\ee
where in the left hand side we consider the integral over the whole chain of the matrices, including $\Phi$, associated with the tau-function. If $O_m$ contains a factor $O_+$ or $O_-$, then according to \eqref{Rassch} the content product $r^{(m)}_{\lambda_m}$
is proportional to the ratio of the Schur functions $s_\lambda$ on a certain locus. 
Examples of the superintegrability rule are considered in Section \ref{S5}.

Similarly we have a general superintegrability rule associated with the variables ${\bf t_0}$ instead of ${\bf t_{m+1}}$.

\subsection{$W$-operators}\label{S4.4}

Let 
\be
O_{(p)}=\prod_{j=1}^p O_+(u_j)
\ee
 be a product of $p$ group operators $O_+(u)$ associated with the elementary weight generating functions $G^+(z)=1+uz$. Then for positive $k$ we consider the operators \cite{WLZZ,Oconj}
\begin{equation}
\begin{split}
W^{(p)}_{k}&=O_{(p)}^{-1} J_{k} O_{(p)},\\
W^{(p)}_{-k}&=O_{(p)} J_{-k} O_{(p)}^{-1}.
\end{split}
\end{equation}
These operators depend on the parameters $u_j$.
Operators $W^{(1)}_{-1}$ and $W^{(2)}_{-1}$ were constructed by Zograf  \cite{Zograf}, and for higher $p$ operators $W^{(p)}_{-1}$ were described in \cite{AMMN} and \cite[Section 5.2]{Ferm}.
For finite $p$ these operators are of finite degree in the bosonic operators $J_k$ and can be described explicitly by the nested commutation relations \cite{WLZZ,MMCH}. By construction, operators $W^{(p)}_{k}$ belong to $\gl(\infty)$
algebra of symmetries of the KP hierarchy.

For a given diagonal group element $O_{(p)}$ operators $W^{(p)}_{k}$ with positive and negative $k$ generate two abelian algebras,
\be
\left[W^{(p)}_{j},W^{(p)}_{k}\right]=0 \quad  \quad  \forall j,k>0 \quad \mathrm{or} \quad  j,k<0
\ee
in the same way as the bosonic operators $J_k$.
However, operators with different signs, in general, do not generate the Heisenberg algebra.

Equation \eqref{JtoJ} allows us to represent $W^{(p)}_{k}$ as operators, acting on the functions of ${\bf t}$ variables,
\begin{equation}
\begin{split}
\lvacn \gamma_+({\bf t }) \,  W^{(p)}_{k}= \widehat{W}^{(p)}_{k} ({\bf t},n) \cdot \lvacn \gamma_+({\bf t }).
\end{split}
\end{equation}
The operators $\widehat{W}^{(p)}_{k} ({\bf t},n) $ are differential operators of finite degree and depend on $n$. For $k>0$ we have
\begin{equation}\label{cajW}
\begin{split}
\widehat{W}^{(p)}_{k}({\bf t},n) &=\widehat{O}_{(p)}^{-1}({\bf t},n)  \, \frac{\p}{ \p t_k} \, \widehat{O}_{(p)}({\bf t},n) ,\\
\widehat{W}^{(p)}_{-k}({\bf t},n) &=\widehat{O}_{(p)}({\bf t},n) \, k t_k \, \widehat{O}_{(p)}^{-1}({\bf t},n) .
\end{split} 
\end{equation}

\begin{remark}\label{R4.6}
There is also an equivalent description,
\be
W^{(p)}_{k}\, \gamma_-({\bf t }) \rvacn=  \widehat{W}^{(p)}_{-k}({\bf t},n) \cdot {\gamma_-({\bf t })} \rvacn.
\ee
\end{remark}

Hence, if $O_m=O_{(p)}$ for a finite $p$ and some $u_j$, and $\sigma_m=-$, then
\begin{equation}
\begin{split}\label{Woper}
\lvacn {\gamma_+({\bf t_{m+1}})} \, O_{m}  \, {\gamma_{\sigma_m}({\bf t_{m}})}  \, O_{m-1} &=  \lvacn {\gamma_+({\bf t_{m+1}})} \, O_{m} \, {\gamma_-({\bf t_{m}})}\,O_{m}^{-1} O_{m}   O_{m-1}\\
&  =  \lvacn {\gamma_+({\bf t_{m+1}})}  \, e^{\sum_{k=1}^\infty t_{m,k} W^{(p)}_{-k} }\, O_{m}  O_{m-1}\\
&= e^{\sum_{k=1}^\infty t_{m,k} \widehat{W}^{(p)}_{-k}({\bf t_{m+1}},n)} \cdot  \lvacn {\gamma_+({\bf t_{m+1}})}\,  O_{m}  O_{m-1}.
\end{split}
\end{equation}
As a result, in this case for the nested hypergeometric tau-function \eqref{taudef} we have
\be
\tau_n^{m,\sigma}({\bf t_{m+1}},\dots,{\bf t_{0}})= e^{\sum_{k=1}^\infty t_{m,k} \widehat{W}^{(p)}_{-k}({\bf t_{m+1}},n)} \cdot \tau_n^{m-1,\sigma'}({\bf t_{m+1}},{\bf t_{m-1}},\dots,{\bf t_0}),
\ee
where 
\be\label{m-1t}
\tau_n^{m-1,\sigma'}({\bf t_{m+1}},{\bf t_{m-1}},\dots,{\bf t_0})= \lvacn {\gamma_+({\bf t_{m+1}})} \, O_{m}  O_{m-1} \, {\gamma_{\sigma_{m-1}}({\bf t_{m-1}})}  \dots   {\gamma_{\sigma_1}({\bf  t_{1}})}
O_{0}  \, {\gamma_-({\bf  t_{0}})} \rvacn
\ee
is also a nested hypergeometric tau-function.

Similarly, if $O_m=O_{(p)}^{-1}$ for a finite $p$ and some $u_j$, and $\sigma_m=+$, we have
\begin{equation}
\begin{split}\label{Woper-}
\lvacn {\gamma_+({\bf t_{m+1}})} \,O_{m}\, {\gamma_{\sigma_m}({\bf t_{m}})}  \, O_{m-1} &=  \lvacn {\gamma_+({\bf t_{m+1}})} \, O_{m} {\gamma_+({\bf t_{m}})}\, O_{m}^{-1} O_{m}   O_{m-1}\\
&  =  \lvacn {\gamma_+({\bf t_{m+1}})}  \, e^{\sum_{k=1}^\infty t_{m,k} W^{(p)}_{k} }\,O_{m}  O_{m-1}\\
&= e^{\sum_{k=1}^\infty t_{m,k} \widehat{W}^{(p)}_{k}({\bf t_{m+1}},n)} \cdot  \lvacn {\gamma_+({\bf t_{m+1}})} \, O_{m}  O_{m-1}
\end{split}
\end{equation}
and
\be
\tau_n^{m,\sigma}({\bf t_{m+1}},\dots,{\bf t_{0}})= e^{\sum_{k=1}^\infty t_{m,k} \widehat{W}^{(p)}_{k}({\bf t_{m+1}},n)} \cdot \tau_n^{m-1,\sigma'}({\bf t_{m+1}},{\bf t_{m-1}},\dots,{\bf t_0}),
\ee
where $\tau_n^{m-1,\sigma'}$ is given by \eqref{m-1t}.

Consider a nested hypergeometric function of the following form
 \begin{multline}\label{sss}
\tau_n^{m,\sigma}=\lvacn {\gamma_+({\bf t_{m+1}})} \, O_{(p_m)}^{-\sigma_m} \, {\gamma_{\sigma_{m}}({\bf t_{m}})} \, O_{(p_m)}^{\sigma_m} O_{(p_{m-1})}^{-\sigma_{m-1}} \, {\gamma_{\sigma_{m-1}}({\bf t_{m-1}})} \, O_{(p_{m-1})}^{\sigma_{m-1}}\dots \\
 O_{(p_1)}^{-\sigma_1} \, {\gamma_{\sigma_1}({\bf  t_{1}})} \, O_{(p_1)}^{\sigma_1} O_{(p_0)} \,  {\gamma_-({\bf  t_{0}})} \rvacn,
\end{multline}
where $O_{(p_k)}=\prod_{j=1}^{p_k} O_+(u_j^{(k)})$ and we denote $O^\pm:=O^{\pm1}$. In particular, the diagonal group elements in this case satisfy the constraint $\prod_{j=1}^m O_j=O_{(p_0)}$.
The tau-function can be described in terms of the operators $W^{(p_i)}_k$, 
 \begin{multline}
\tau_n^{m,\sigma}= \lvacn {\gamma_+({\bf t_{m+1}})} e^{\sum_{k=1}^\infty t_{m,k} W^{(p_m)}_{\sigma_m k} }  e^{\sum_{k=1}^\infty t_{m-1,k} W^{(p_{m-1})}_{\sigma_{m-1} k} } \dots\\
 \dots  e^{\sum_{k=1}^\infty t_{1,k} W^{(p_1)}_{\sigma_1 k}}  e^{\sum_{k=1}^\infty t_{0,k} W^{(p_0)}_{- k}} O_{(p_0)} \rvacn,
\end{multline}
and, applying \eqref{Oact} we get
 \begin{multline}
\tau_n^{m,\sigma}= c_n^{(0)} \lvacn {\gamma_+({\bf t_{m+1}})} e^{\sum_{k=1}^\infty t_{m,k} W^{(p_m)}_{\sigma_m k} }  e^{\sum_{k=1}^\infty t_{m-1,k} W^{(p_{m-1})}_{\sigma_{m-1} k} } \dots\\
 \dots  e^{\sum_{k=1}^\infty t_{1,k} W^{(p_1)}_{\sigma_1 k}}  e^{\sum_{k=1}^\infty t_{0,k} W^{(p_0)}_{- k}}  \rvacn.
\end{multline}

By \eqref{Woper} and \eqref{Woper-}, this vacuum expectation value can also be described by the action of the cut-and-join operators $\widehat{W}$ on the trivial function.
\begin{proposition}\label{prop4.2} A nested hypergeometric tau-function \eqref{sss} is given by a recursive action of the exponential operators,
\begin{multline}\label{prtv}
\tau_n^{m,\sigma}=c_n^{(0)}\, e^{\sum_{k=1}^\infty t_{m,k} \widehat{W}^{(p_m)}_{\sigma_m k} ({\bf t_{m+1}},n)}  \, e^{\sum_{k=1}^\infty t_{m-1,k} \widehat{W}^{(p_{m-1})}_{\sigma_{m-1} k} ({\bf t_{m+1}},n)} \dots\\
\dots   e^{\sum_{k=1}^\infty t_{1,k} \widehat{W}^{(p_1)}_{\sigma_1 k}({\bf t_{m+1}},n) } e^{\sum_{k=1}^\infty t_{0,k} \widehat{W}^{(p_0)}_{- k}({\bf t_{m+1}},n) } \cdot 1.
\end{multline}
\end{proposition}
Note that the operators $\widehat{W}^{(p_j)}_{\sigma_j k} ({\bf t_{m+1}},n)$  and  $\widehat{W}^{(p_\ell)}_{\sigma_\ell i} ({\bf t_{m+1}},n)$ for $j \neq \ell$, in general, do not commute with each other. 

The nested hypergeometric tau-function \eqref{sss} has a skew Schur function expansion
\be
\tau_n^{m,\sigma}=\prod_{j=0}^m c_n^{(j)}\sum_{\lambda_0,\dots,\lambda_m} s_{\lambda_m}({\bf t_{m+1}})  \prod_{i=1}^m \left( \frac{r^{(i)}_{\lambda_{i-1},n}}{r^{(i)}_{\lambda_i,n}}  s_{(\lambda_{i-1}/\lambda_i)^{\sigma_i}}({\bf t_{i}})\right) \, r^{(0)}_{\lambda_0,n} \, s_{\lambda_0}({\bf t_0}),
\ee
where $r^{(i)}_{\lambda,n}=\prod_{j=1}^{p_i} r^{\sigma_i}_{\lambda,n}(u_j^{(i)})$.

If we consider a specialisation of \eqref{sss} with $O_{(p_0)}=1$ (in this case $\prod_{j=1}^m O_j=1$),
 then \eqref{prtv} reduces to 
 \begin{multline}\label{cajo}
\tau_n^{m,\sigma}=e^{\sum_{k=1}^\infty t_{m,k} \widehat{W}^{(p_m)}_{\sigma_m k} ({\bf t_{m+1}},n)}  e^{\sum_{k=1}^\infty t_{m-1,k} \widehat{W}^{(p_{m-1})}_{\sigma_{m-1} k} ({\bf t_{m+1}},n)} \dots\\
\dots   e^{\sum_{k=1}^\infty t_{1,k} \widehat{W}^{(p_1)}_{\sigma_1 k}({\bf t_{m+1}},n) }\cdot e^{\sum_{k=1}^\infty k\, t_{m+1,k}\, t_{0,k}}.
\end{multline}
In this case, a dual form of the nested hypergeometric tau-function \eqref{sss}, 
\begin{multline}
\tau_n^{m,\sigma}=\lvacn {\gamma_+({\bf t_{0}})} \,O_{(p_1)}^{\sigma_1} \,{\gamma_{-\sigma_1}({\bf  t_{1}})} \, O_{(p_1)}^{-\sigma_1} \dots\\
\dots O_{(p_{m-1})}^{\sigma_{m-1}} \,{\gamma_{-\sigma_{m-1}}({\bf t_{m-1}})}\, O_{(p_{m-1})}^{-\sigma_{m-1}}\,
O_{(p_m)}^{\sigma_m}\, {\gamma_{-\sigma_{m}}({\bf t_{m}})} \,O_{(p_m)}^{-\sigma_m} \,
  {\gamma_-({\bf  t_{m+1}})} \rvacn,
\end{multline}
 immediately implies the cut-and-join description dual to \eqref{cajo},
\begin{multline}\label{dualW}
\tau_n^{m,\sigma}=  e^{\sum_{k=1}^\infty t_{1,k} \widehat{W}^{(p_1)}_{-\sigma_1 k}({\bf t_{0}},n) }\dots\\
\dots   e^{\sum_{k=1}^\infty t_{m-1,k} \widehat{W}^{(p_{m-1})}_{-\sigma_{m-1} k} ({\bf t_{0}},n)} \,
e^{\sum_{k=1}^\infty t_{m,k} \widehat{W}^{(p_m)}_{-\sigma_m k} ({\bf t_{0}},n)}   \cdot e^{\sum_{k=1}^\infty k \,t_{m+1,k} \, t_{0,k}}.
\end{multline}
\begin{remark}
One can also apply Remark \ref{R4.6} to represent the tau-function \eqref{cajo} as a combination of the operators acting both on ${\bf t_{m+1}}$ and ${\bf t_0}$. Namely, for any $j$ such that $m\geq j\geq 0$ we have
 \begin{multline}
\tau_n^{m,\sigma}=e^{\sum_{k=1}^\infty t_{m,k} \widehat{W}^{(p_m)}_{\sigma_m k} ({\bf t_{m+1}},n)} \dots e^{\sum_{k=1}^\infty t_{j+1,k} \widehat{W}^{(p_{j+1})}_{\sigma_{j+1} k} ({\bf t_{m+1}},n)}\\
e^{\sum_{k=1}^\infty t_{1,k} \widehat{W}^{(p_1)}_{-\sigma_1 k}({\bf t_{0}},n) }\dots e^{\sum_{k=1}^\infty t_{j,k} \widehat{W}^{(p_j)}_{-\sigma_j k}({\bf t_{0}},n) } \cdot e^{\sum_{k=1}^\infty k\, t_{m+1,k}\, t_{0,k}}.
\end{multline}
The simplest case when such a description with mixed $\widehat W$-operators appears corresponds to $m=2$, see \eqref{mixedW}.
\end{remark}

To describe some nested hypergeometric tau-functions with the weight generating functions \eqref{Gcovered} one may need to combine the cut-and-join descriptions of two types, $\widehat{W}^{(p)}_k$ operators and operators $\widehat{O}$ from Section \ref{S3.2}. In particular, the weight generating function $G^{\exp}$ is described by the original cut-and-join operator \cite{GJ}.

\section{Examples}\label{S5}

In this section, we discuss various examples of the nested hypergeometric tau-functions, their skew Schur function expansions, description by matrix models and cut-and-join operators, and relation to the enumerative geometry invariants.

\subsection{Hypergeometric tau-functions ($m=0$)}

For $m=0$ a nested hypergeometric tau-function reduces to a hypergeometric (or Orlov--Scherbin) tau-function
\begin{equation}
\begin{split}\label{Hygo}
\tau_n^{0,\emptyset}({\bf t_{1}},{\bf t_0})
&=\lvacn {\gamma_+({\bf t_{1}})} \, O_{0} \, {\gamma_-({\bf  t_{0}})} \rvacn\\
&=c_n^{0}\sum_{\lambda_0}  \,s_{\lambda_0}({\bf t_1})\, r^{(0)}_{\lambda_0,n}\, s_{\lambda_0}({\bf t_0}),
\end{split}
\end{equation}
first considered in \cite{KMMM} and further investigated in \cite{OS}.
Matrix models for the hypergeometric tau-functions, associated with the weighted Hurwitz numbers are discussed in \cite{ZJZ,Orlov,O2,Eynard,MMRP, AMMN,GGPN,AC,Orlovchain}. 

Applying the general construction of Section \ref{S4.2} one can derive a multi-matrix model description for the generating function of the weighed Hurwitz numbers for any weight generating function of the form \eqref{Gcovered}.  From Proposition \ref{prop4.1} we immediately get a description of the hypergeometric tau-functions by a chain of matrix models.
\begin{corollary}
A hypergeometric tau-function \eqref{Hygo} with diagonal group operator $O_0$ associated with the weight generating function \eqref{Gcovered} is described by a chain of
$\ell$ integrals \eqref{sumpl}, $n$ integrals \eqref{Uele}, and, if $w\neq 0$, one integral \eqref{log} with a pairwise interaction and arbitrary order of the elementary integrals. Moreover, these multi-matrix integrals can be reduced to the eigenvalue integrals. 
\end{corollary}

Let us consider an example. The weight generating function 
\be\label{weightppm1}
G(z)=\frac{1+N_3^{-1}z}{(1+N_1^{-1}z)(1+N_2^{-1}z)}
\ee
corresponds to the content product
\be\label{weightppm}
r_\lambda=r_\lambda^{+}(N_3^{-1})r_\lambda^{-}(N_2^{-1})r_\lambda^{-}(N_1^{-1}).
\ee
According to Proposition \ref{prop4.1} in this case the tau-function can be described by a chain of two unitary matrix integrals \eqref{Uele} and one complex matrix integral \eqref{sumpl}
\begin{multline}\label{chain3}
\sum_{\ell(\lambda)\leq \min (N_1,N_2,N_3)} s_\lambda({\bf t_1})\,  r_\lambda \, s_\lambda({\bf t_0}) = \int_\mathfrak{C} \left[d\mu_\mathfrak{C} ({Z}_3)\right]  \int_\mathfrak{U^2}\left[d\mu_\mathfrak{U}{(U_2)}\right]  \int_\mathfrak{U^2}\left[d\mu_\mathfrak{U}{(U_1)}\right]\times\\
\times \exp\left( \sum_{k=1}^\infty t_{1,k} \Tr Z_3^k+\frac{1}{k} \Tr (Z_3^\dagger)^k \Tr \tilde{U}_2^k +\frac{1}{k} \Tr (U_2^\dagger)^k \Tr \tilde{U}_1^k+ t_{0,k} \Tr (U_1^\dagger )^k\right).
\end{multline}
Here the lower index of the matrix indicates its size, for instance, $U_2$ and $\tilde U_2$ are $N_2\times N_2$ unitary matrices. Indeed, let us apply equations \eqref{Uele} and \eqref{elesu} to the right hand side of \eqref{chain3},
\begin{equation}
\begin{split}\label{derivv}
&\int_\mathfrak{C} \left[d\mu_\mathfrak{C} ({Z}_3)\right]  \int_\mathfrak{U^2}\left[d\mu_\mathfrak{U}{(U_2)}\right]  \int_\mathfrak{U^2}\left[d\mu_\mathfrak{U}{(U_1)}\right]\times\\
&\times \exp\left( \sum_{k=1}^\infty t_{1,k} \Tr Z_3^k+\frac{1}{k} \Tr (Z_3^\dagger)^k \Tr \tilde{U}_2^k +\frac{1}{k} \Tr (U_2^\dagger)^k \Tr \tilde{U}_1^k+ t_{0,k} \Tr (U_1^\dagger )^k\right)\\
&=\int_\mathfrak{C} \left[d\mu_\mathfrak{C} ({Z}_3)\right]  \int_\mathfrak{U^2}\left[d\mu_\mathfrak{U}{(U_2)}\right]  \times\\
&\times 
 \sum_{\ell(\lambda)\leq N_1} s_\lambda(U_2^\dagger) r_\lambda^{-}(N_1^{-1}) s_\lambda({\bf t_0}) \exp\left( \sum_{k=1}^\infty t_{1,k} \Tr Z_3^k+\frac{1}{k} \Tr (Z_3^\dagger)^k \Tr \tilde{U}_2^k\right) \\
& =\int_\mathfrak{C} \left[d\mu_\mathfrak{C} ({Z}_3)\right]  
 \sum_{\ell(\lambda)\leq  \min (N_1,N_2)} s_\lambda(Z_3^\dagger) r_\lambda^{-}(N_2^{-1})  r_\lambda^{-}(N_1^{-1}) s_\lambda({\bf t_0}) \exp\left( \sum_{k=1}^\infty t_{1,k} \Tr Z_3^k\right)\\
 &= \sum_{\ell(\lambda)\leq  \min (N_1,N_2,N_3)} s_\lambda({\bf t_1}) r_\lambda^{+}(N_3^{-1}) r_\lambda^{-}(N_2^{-1})   r_\lambda^{-}(N_1^{-1}) s_\lambda({\bf t_0})  .
\end{split}
\end{equation}
Moreover, from this example it is clear that the order of the elements of the chain of matrix integrals is arbitrary, say, function \eqref{chain3} has as well another matrix integral description:
\begin{multline}
\sum_{\ell(\lambda)\leq \min (N_1,N_2,N_3)} s_\lambda({\bf t_1})\,  r_\lambda \, s_\lambda({\bf t_0}) = \int_\mathfrak{U^2}\left[d\mu_\mathfrak{U}{(U_2)}\right]  \int_\mathfrak{C} \left[d\mu_\mathfrak{C} ({Z}_3)\right]   \int_\mathfrak{U^2}\left[d\mu_\mathfrak{U}{(U_1)}\right]\times\\
\times \exp\left( \sum_{k=1}^\infty t_{1,k} \Tr \tilde{U}_2^k+\frac{1}{k} \Tr ({U}_2^\dagger)^k \Tr Z_3^k +\frac{1}{k} \Tr (Z_3^\dagger)^k \Tr \tilde{U}_1^k+ t_{0,k} \Tr (U_1^\dagger )^k\right).
\end{multline}

For any two matrices $\Phi_1$ and $\Phi_2$ we have
\be
\exp\left( \sum_{k=1}^\infty  \frac{1}{k}\Tr \Phi_1^k \, \Tr \Phi_2^k\right) =\frac{1}{\det \left(I_1\otimes I_2- \Phi_1\otimes \Phi_2\right)},
\ee
where $I_1$ and $I_2$ are the identity matrices of the  size same as $\Phi_1$ and $\Phi_2$. Then the matrix integral \eqref{chain3} can be represented as
\be\label{mmppm}
 \int_\mathfrak{C} \left[d\mu_\mathfrak{C} ({Z}_3)\right]  \int_\mathfrak{U^2} \left[d\mu_\mathfrak{U}{(U_2)}\right]  \int_\mathfrak{U^2} \left[d\mu_\mathfrak{U}{(U_1)}\right]\frac{\displaystyle{
\exp\left( \sum_{k=1}^\infty  t_{1,k} \Tr Z_3^k + t_{0,k} \Tr (U_1^\dagger )^k\right)}}{\det\left(I_1\otimes I_2- \tilde{U}_1\otimes U_2^\dagger\right)\det\left(I_2\otimes I_3-  \tilde{U}_2\otimes Z_3^\dagger\right)}.
\ee

Here we see the general property of the constructed chains of the matrix integrals -- all interactions are pairwise of the form $\exp(N \, \Tr \Phi_1\Phi_2)$ (in this example they are hidden in the expressions for the integration measures $[d \mu_\bullet]$) or $\frac{1}{\det \left(I_1\otimes I_2- \Phi_1\otimes \Phi_2\right)}$, and every matrix (and its conjugate for normal and complex matrix integrals) interacts with two other matrices of integration, except for the matrices corresponding to the ends of the chain, which interact with only one matrix each. These matrices corresponding to the ends of the chain (in our example they are $U_1$ and $Z_3$) couple to the times of the integrable hierarchy ${\bf t_0}$ and ${\bf t_{m+1}}$.

\begin{remark}
Note that the diagonal operators $O$ commute with each other, therefore the order of the matrix integrals for the elementary factors of the diagonal group element $O$ can be arbitrary. This is also correct for the elementary factors of the  operators $O_j$ in the matrix models of the nested hypergeometric tau-functions.
\end{remark}

\begin{remark}
Let us note again that the general construction of Section \ref{S4.2} provides a universal matrix integral expression suitable for all nested hypergeometric tau-functions \eqref{taudef} with the weight generating functions \eqref{Gcovered}. The price we pay for this universality is that the obtained multi-matrix integral, in general, is not the simplest possible one.  There are equivalent representations, for instance, the tau-function \eqref{chain3} in the Miwa parametrization can be described by a chain of a two-matrix model and two unitary integrals,
\begin{multline}\label{altdoub}
\sum_{\ell(\lambda)\leq \min (N_1,N_2,N_3)} s_\lambda(A)\, r_\lambda \, s_\lambda(B) \\
=\int_\mathfrak{U} \left[d{U_1}\right]  \int_{\mathfrak{H}^2} \left[d{X}\right] \left[d{Y}\right]  \int_\mathfrak{U} \left[d{U_2}\right] 
\frac{\exp\left(N_1 \Tr{U_1^\dagger A} +N_2 \Tr{U_2^\dagger B} +N_3\sqrt{-1} \Tr {XY} \right)}{\det\left(I_1\otimes I_3- {U}_1\otimes X\right)\det\left(I_2\otimes I_3- {U}_2\otimes Y\right)}.
\end{multline}
Here one integrates over the unitary $N_1\times N_1$ and $N_2\times N_2$ matrices $U_1$ and $U_2$ respectively and over the $N_3\times N_3$ Hermitian matrices $X$ and $Y$. Indeed, applying \eqref{unit2} twice we have
\begin{equation}
\begin{split}
&\int_\mathfrak{U} \left[d{U_1}\right]  \int_{\mathfrak{H}^2} \left[d{X}\right] \left[d{Y}\right]  \int_\mathfrak{U} \left[d{U_2}\right] 
\frac{\exp\left(N_1 \Tr{U_1^\dagger A} +N_2 \Tr{U_2^\dagger B} +N_3\sqrt{-1} \Tr {XY} \right)}{\det\left(I_1\otimes I_3- {U}_1\otimes X\right)\det\left(I_2\otimes I_3- {U}_2\otimes Y^\dagger\right)}\\
&= \int_{\mathfrak{H}^2} \left[d{X}\right] \left[d{Y}\right] \sum_{\ell(\lambda)\leq N_1} s_\lambda(A) r_\lambda^{-}(N_1^{-1}) s_\lambda(X)  \sum_{\ell(\mu)\leq N_2} s_\mu(B) r_\mu^{-}(N_2^{-1}) s_\mu(Y)
\exp\left(N_3\sqrt{-1} \Tr {XY} \right),
\end{split}
\end{equation}
and \eqref{altdoub} follows from the Cauchy--Littlewood formula expansion of \eqref{double}.
\end{remark}

A hypergeometric tau-function can be described by the cut-and-join operators of Section \ref{S3.2}, 
\be\label{Ohyp}
\tau_n^{0,\emptyset}({\bf t_{1}},{\bf t_0})=\widehat{O}_0({\bf t_1},n)\cdot \exp\left(\sum_{k=1}^\infty k \, t_{0,k} \, t_{1,k}\right)
\ee
and
\be\label{Ohyp1}
\tau_n^{0,\emptyset}({\bf t_{1}},{\bf t_0})=\widehat{O}_0({\bf t_0},n)\cdot \exp\left(\sum_{k=1}^\infty k \, t_{0,k} \, t_{1,k}\right).
\ee

For a polynomial weight generating function $G(z)= \prod_{i=1}^p (1+u_i z) $ our general construction yields a chain of the complex matrix models, which is equivalent to a chain of the two-matrix integrals constructed in
\cite{AMMN}. In this case, the tau-function is the simplest representative of the family \eqref{sss}, and, according to Proposition \ref{prop4.2} it can be described \cite{Adr,Oconj,MMCH} by the cut-and-join operators \eqref{cajW}, 
\begin{equation}
\begin{split}\label{hyperm}
\tau_n^{0,\emptyset}({\bf t_{1}},{\bf t_0})
&=\lvacn {\gamma_+({\bf t_{1}})} \, O_{(p)} \, {\gamma_-({\bf  t_{0}})} \rvacn \\
&= c_n^{(0)} \lvacn {\gamma_+({\bf t_{1}})} \exp\left(\sum_{k=1}^\infty t_{0,k} W^{(p)}_{k} \right) \rvacn\\
&= c_n^{(0)}  \exp\left(\sum_{k=1}^\infty t_{0,k} \widehat{W}^{(p)}_{k}({\bf t_1},n) \right) \cdot 1,
\end{split}
\end{equation}
or, equivalently
\be
\tau_n^{0,\emptyset}({\bf t_{1}},{\bf t_0})= c_n^{(0)}  \exp\left(\sum_{k=1}^\infty t_{1,k} \widehat{W}^{(p)}_{k}({\bf t_0},n) \right) \cdot 1.
\ee 

In particular, the generating function of  {\em dessins d'enfant} (or hypermaps, or the strictly monotone Hurwitz numbers, or  the complex matrix model \eqref{sumpl}, or the two-matrix model \eqref{double}) corresponds to $O_0=O_+(\hbar)$ and $n=0$. In this case we have \cite{Adr}
\begin{equation}
\begin{split}
\tau_0^{0,\emptyset}({\bf t_{1}},{\bf t_0})&=\exp\left(\sum_{k=1}^\infty t_{0,k} \widehat{W}^{(1)}_{k}({\bf t_1},0) \right) \cdot 1\\
&=\exp\left(\sum_{k=1}^\infty t_{1,k} \widehat{W}^{(1)}_{k}({\bf t_0},0) \right) \cdot 1.
\end{split}
\end{equation}
Further simplification to the case of {\em maps} (or Gaussian Hermitian matrix model) corresponds to $t_{0,k}=\delta_{k,2}/2\hbar$, in this case we have \cite{MS}
\be\label{smW}
\tau_0^{0,\emptyset}({\bf t_{1}})=  \exp\left(\frac{1}{2\hbar} \widehat{W}^{(1)}_{2}({\bf t_1},0) \right) \cdot 1.
\ee
\begin{remark}
Cut-and-join description and the Schur function expansion for the so-called {\em stuffed maps} can be considered in a similar way, see Section \ref{S6}.
\end{remark}



All matrix integrals obtained this way can be interpreted in the framework of  superintegrability as compositions of the elementary cases \eqref{elesu}. For example, for the weight generating function \eqref{weightppm1} from the matrix integral \eqref{mmppm}
we have a relation
\begin{multline}
 r_\lambda \, s_\lambda({\bf t_0}) =\int_\mathfrak{C} \left[d\mu_\mathfrak{C} ({Z}_3)\right]  \int_\mathfrak{U^2} \left[d\mu_\mathfrak{U}{(U_2)}\right]  \int_\mathfrak{U^2} \left[d\mu_\mathfrak{U}{(U_1)}\right]\times \\
\times s_\lambda(Z_3) \frac{\displaystyle{ 
\exp\left( \sum_{k=1}^\infty t_{0,k} \Tr (U_1^\dagger )^k\right)}}{\det\left(I_1\otimes I_2- \tilde{U}_1\otimes U_2^\dagger\right)\det\left(I_2\otimes I_3-  \tilde{U}_2\otimes Z_3^\dagger\right)}
\end{multline}
and its dual
\begin{multline}
 r_\lambda \, s_\lambda({\bf t_1}) =\int_\mathfrak{C} \left[d\mu_\mathfrak{C} ({Z}_3)\right]  \int_\mathfrak{U^2} \left[d\mu_\mathfrak{U}{(U_2)}\right]  \int_\mathfrak{U^2} \left[d\mu_\mathfrak{U}{(U_1)}\right]\times \\
\times s_\lambda(U_1^\dagger) \frac{\displaystyle{ 
\exp\left( \sum_{k=1}^\infty  t_{1,k} \Tr Z_3^k \right)}}{\det\left(I_1\otimes I_2- \tilde{U}_1\otimes U_2^\dagger\right)\det\left(I_2\otimes I_3-  \tilde{U}_2\otimes Z_3^\dagger\right)}.
\end{multline}

\subsection{Skew hypergeometric tau-functions ($m=1$)}\label{S5.2}
In this section, we consider the nested hypergeometric tau-functions with $m=1$ in more detail. For this case there are two different options, namely $\sigma=+$ and $\sigma=-$, but because of duality \eqref{ptom} they are equivalent to each other, therefore, it is enough to consider only the $+$ sign,
\begin{equation}
\begin{split}\label{Q}
 \tau_n^{1,+}&=\lvacn {\gamma_{+}({\bf  t_{2}})} \, O_{1} \, {\gamma_{+}({\bf  t_{1}})} \, O_{0} \, {\gamma_-({\bf  t_{0}})}\rvacn\\
 &= c_n^{(0)}c_n^{(1)}\sum_{\lambda_0,\lambda_1}s_{\lambda_1}({\bf t_2})\, r^{(1)}_{\lambda_1,n}\, s_{\lambda_0/\lambda_1}({\bf t_1})\, r^{(0)}_{\lambda_0,n}\,  s_{\lambda_0}({\bf t_0}).
\end{split}
\end{equation}
However, below we will use the symmetry between the tau-functions labeled by different signs to investigate the relation between different versions of the cut-and-join descriptions. 

 If either $O_0=1$ or $O_1=1$, then the nested hypergeometric tau-function reduces to the hypergeometric tau-function considered in the previous section. Therefore, we assume that both $O_0\neq 1$ and $O_1\neq 1$. 
 
 The tau-function given by the last line of \eqref{Q} was recently introduced and investigated by Mironov, Mishnyakov, Morozov, Popolitov, Wang, and Zhao \cite{Oconj}. They call it the {\em skew hypergeometric tau-function}.
Developing the approach of \cite{WLZZ,Adr},  for a particular choice of the diagonal group elements they also found the matrix model and the cut-and-join description, see below. In this section we will apply the general construction to re-derive and generalize the description of \cite{Oconj}.

The case $O_{0}=O_{1}^{-1}=O_+(\hbar)$  can be identified with the generating function of interesting enumerative geometry invariants.
Namely, it provides the generating function of the fully simple maps, recently introduced and investigated by Borot and Garcia-Failde
\cite{BGF}, and developed, in particular, in \cite{BCDG,BCGF}. We refer to these papers for the definition of the fully simple maps, and consider below only their generating function. 
In \cite{BDKS} Bychkov, Dunin-Barkowski, Kazarian, and  Shadrin identify this generating function with the following vacuum expectation value
\be\label{FSS}
Z=\lvac {\gamma_+({\bf t_2})}\, O_+^{-1}(\hbar)\, {\gamma_+(\hbar^{-1}{\bf t_1})}\, O_+(\hbar) \, e^{\frac{1}{2 \hbar}J_{-2}} \rvac,
\ee
where the group element $O_+(\hbar)$ corresponds to the weight generating function $G_+(z)=1+\hbar z$.

Comparing \eqref{FSS} with \eqref{Q} we see that $Z$ is a nested hypergeometric tau-function for $n=0$, $m=1$,  $O_{0}=O_{1}^{-1}=O_+(\hbar)$, and $t_{0,k}=\delta_{k,2}/(2\hbar)$.
Therefore, we have an immediate corollary of Theorem \ref{T1}.
\begin{corollary}
The generating function for the fully simple maps has the following skew Schur function expansion
\be
Z=\sum_{\lambda_0,\lambda_1} \frac{r^+_{\lambda_0}(\hbar)}{r^+_{\lambda_1}(\hbar)} s_{\lambda_1}({\bf t_2}) s_{\lambda_0/\lambda_1}({\bf t_1}/\hbar)s_{\lambda_0}(\delta_{k,2}/2\hbar).
\ee
\end{corollary}

In \cite[Section 4.1]{BDKS} the authors introduce the following  generalization of the generating function \eqref{FSS}
\be\label{Zvee}
Z^{\vee}=\lvac {\gamma_+({\bf t_2})} \, O^{-1} \, {\gamma_+(\hbar^{-1}{\bf t_1})} \, O \, {\gamma_-(\hbar^{-1}{\bf t_0})} \rvac,
\ee
where $O$ is an arbitrary diagonal group element \eqref{diagge}. The enumerative geometry interpretation of this generating function is not clear yet. Please, note that this function is not dual to \eqref{FSS}, but
is a generalization of that. 

This is a nested hypergeometric tau-function with $m=1$.
Again, for this more general case we have an immediate corollary of Theorem \ref{T1}.
\begin{corollary}
The generating function \eqref{Zvee} has the following skew Schur function expansion
\be\label{Zvee1}
Z^{\vee}=\sum_{\lambda_0,\lambda_1} \frac{r_{\lambda_0}}{r_{\lambda_1}} s_{\lambda_1}({\bf t_2}) s_{\lambda_0/\lambda_1}({\bf t_1}/\hbar)s_{\lambda_0}({\bf t_0}/\hbar),
\ee
where $r$ is the content product for the operator $O$.
\end{corollary}

If operators $O_0$ and $O_1$ in \eqref{Q} describe the weight generating functions \eqref{Gcovered}, then the construction of Section \ref{S4.2} immediately gives us a multi-matrix model for the tau-function $ \tau_0^{1,+}$. Let us consider an example associated with the fully simple hypermaps, introduced in \cite{BCDG}, which corresponds to $O_{0}=O_{1}^{-1}=O_+(\hbar)$:
\be\label{fsh}
\lvac {\gamma_+({\bf t_2})}\, O_+^{-1}(\hbar)\, {\gamma_+(\hbar^{-1}{\bf t_1})}\, O_+(\hbar) \, {\gamma_-(\hbar^{-1}{\bf t_0})} \rvac.
\ee
This description of the generating function of fully simple hypermaps follows from their relation to ordinary hypermaps proven in \cite{BCDG} in exactly the same way as for simple maps in \cite{BDKS}. We immediately conclude that the generating function of fully simple hypermaps is a tau-function of 2D Toda lattice hierarchy in the variables related to the boundary faces and hyperedges. 

 In this case, if we put $\hbar=N^{-1}$, our general construction yields the following matrix model:
\begin{multline}\label{MM1}
\sum_{\ell(\lambda_0),\ell(\lambda_1)\leq N}  \frac{r^+_{\lambda_0}(N^{-1})}{r^+_{\lambda_1}(N^{-1})} s_{\lambda_1}({\bf t_2}) s_{\lambda_0/\lambda_1}(N{\bf t_1})s_{\lambda_0}(N{\bf t_0})\\
=\int_\mathfrak{U^2} \left[d\mu_\mathfrak{U}{(U)}\right]  \int_\mathfrak{C} \left[d \mu_\mathfrak{C}({Z})\right] \frac{\displaystyle{ \exp\left( \Tr \sum_{k=1}^\infty  t_{2,k} {\tilde U}^k+ N \, t_{1,k}{Z}^k + N\, t_{0,k}{Z^\dagger}^k \right)}}{\det\left(I\otimes I- Z\otimes U^\dagger\right)}.
\end{multline}

Moreover, because of the additional symmetries  in this case the matrix integral can be further simplified. Indeed, comparing \eqref{Uele} with \eqref{unit2} we conclude that
\be\int_\mathfrak{U^2} \left[d\mu_\mathfrak{U}{(U)}\right]  \frac{\displaystyle{ \exp\left( \Tr \sum_{k=1}^\infty  t_{2,k} {\tilde U}^k \right)}}{\det\left(I\otimes I- Z\otimes U^\dagger\right)}=\int_\mathfrak{U} \left[d U\right]  \exp\left(\sum_{k=1}^\infty  t_{2,k} \Tr{U}^k+ N\, \Tr U^\dagger Z\right),
\ee
therefore we can integrate out one of the unitary matrices:
\begin{multline}\label{Simpfs}
\sum_{\ell(\lambda_0),\ell(\lambda_1)\leq N}  \frac{r^+_{\lambda_0}(N^{-1})}{r^+_{\lambda_1}(N^{-1})} s_{\lambda_1}({\bf t_2}) s_{\lambda_0/\lambda_1}(N{\bf t_1})s_{\lambda_0}(N{\bf t_0})\\
=\int_\mathfrak{U} \left[d U\right]  \int_\mathfrak{C} \left[d \mu_\mathfrak{C}({Z})\right] \exp\left(\sum_{k=1}^\infty  t_{2,k} \Tr{U}^k+ N\, \Tr\left( U^\dagger Z+\sum_{k=1}^\infty t_{1,k}{Z}^k +t_{0,k}{Z^\dagger}^k \right)\right).
\end{multline}

This expression can be further simplified if we consider the Miwa parametrization of the variables ${\bf t_2}$, that is, if we put $t_{2,k}=\frac{1}{k}\Tr \Lambda^k$ for a diagonal $N\times N$ matrix $\Lambda$. Namely, in this case comparing \eqref{unit2} with the Harish-Chandra--Itzykson--Zuber matrix integral \eqref{unit} we immediately conclude
\be
\int_\mathfrak{U} \left[d U\right]  \exp\left(\sum_{k=1}^\infty \frac{1}{k} \Tr \Lambda^k \Tr{U}^k+ N\, \Tr U^\dagger Z\right)= \int_\mathfrak{U} \left[d{U}\right] \exp\left(N \Tr({ {U\Lambda U^\dagger Z}}) \right).
\ee
Substituting it into \eqref{Simpfs} we have
\begin{multline}
\sum_{\ell(\lambda_0),\ell(\lambda_1)\leq N}  \frac{r^+_{\lambda_0}(N^{-1})}{r^+_{\lambda_1}(N^{-1})} s_{\lambda_1}(\Lambda) s_{\lambda_0/\lambda_1}(N{\bf t_1})s_{\lambda_0}(N{\bf t_0})\\
=\int_\mathfrak{U} \left[d U\right]  \int_\mathfrak{C} \left[d \mu_\mathfrak{C}({Z})\right] \exp\left( N\, \Tr\left( {U\Lambda U^\dagger Z}+\sum_{k=1}^\infty t_{1,k}{Z}^k +t_{0,k}{Z^\dagger}^k \right)\right).
\end{multline}
The measure $\left[d \mu_\mathfrak{C}({Z})\right]$ as well as the traces $\Tr {Z}^k$, $\Tr{Z^\dagger}^k$ are invariant under a unitary matrix conjugation $Z\mapsto U Z U^\dagger$, therefore
\begin{multline}
\sum_{\ell(\lambda_0),\ell(\lambda_1)\leq N}  \frac{r^+_{\lambda_0}(N^{-1})}{r^+_{\lambda_1}(N^{-1})} s_{\lambda_1}(\Lambda) s_{\lambda_0/\lambda_1}(N{\bf t_1})s_{\lambda_0}(N{\bf t_0})\\
=\int_\mathfrak{U} \left[d U\right]  \int_\mathfrak{C} \left[d \mu_\mathfrak{C}({Z})\right] \exp\left( N\, \Tr\left( {\Lambda  Z}+\sum_{k=1}^\infty t_{1,k}{Z}^k +t_{0,k}{Z^\dagger}^k \right)\right)\\
= \int_\mathfrak{C} \left[d \mu_\mathfrak{C}({Z})\right]\exp\left( N \, \Tr\left(\Lambda Z+ \sum_{k=1}^\infty t_{1,k}{Z}^k+ t_{0,k}{{Z^\dagger}^k}\right)\right).
\end{multline}

 This matrix integral is equivalent to the two-matrix model
\begin{multline}\label{MM4}
\sum_{\ell(\lambda_0),\ell(\lambda_1)\leq N}  \frac{r^+_{\lambda_0}(N^{-1})}{r^+_{\lambda_1}(N^{-1})} s_{\lambda_1}(\Lambda) s_{\lambda_0/\lambda_1}(N{\bf t_1})s_{\lambda_0}(N{\bf t_0})\\
=\int_{\mathfrak{H}^2} \left[d{X}\right]  \left[d{Y}\right]\exp\left(N \, \Tr \left( {\Lambda Y} + \sqrt{-1}{XY} +\sum_{k=1}^\infty t_{1,k}{Y}^k+  t_{0,k}{X^k}\right)\right).
\end{multline}
The last identity was derived in \cite{MMCH}. The matrix models \eqref{MM1}--\eqref{MM4} describe the generating function of fully simple hypermaps \eqref{fsh}.

For $t_{0,k}=\delta_{k,2}/2$  the integral in $X$ is Gaussian and can be computed explicitly, so we get
\begin{multline}
\sum_{\ell(\lambda_0),\ell(\lambda_1)\leq N} \frac{r^+_{\lambda_0}(N^{-1})}{r^+_{\lambda_1}(N^{-1})} s_{\lambda_1}(\Lambda ) s_{\lambda_0/\lambda_1}(N{\bf t_1})s_{\lambda_0}(N\delta_{k,2}/2)
\\=
\int_{\mathfrak{H}} \left[d{Y}\right]\exp\left( N \, \Tr \left( Y\Lambda -  {Y^2}/2 +\sum_{k=1}^\infty t_{1,k}{Y}^k\right) \right).
\end{multline}
The Hermitian matrix model on the right hand side is the model for the generating function of fully simple maps \eqref{FSS}, derived by Borot and Garcia-Failde \cite{BGF}.

In terms of the cut-and-join operators of Section \ref{S3.2} the nested hypergeometric tau-function \eqref{Q} can be related to the hypergeometric tau-function \eqref{Hygo} in two different ways,
\begin{equation}
\begin{split}\label{O1}
 \tau_n^{1,+}&=\lvacn {\gamma_{+}({\bf  t_{2}})} \, O_{1} \, {\gamma_{+}({\bf  t_{1}})} \, O_{0} \, {\gamma_-({\bf  t_{0}})}\rvacn\\
 &=\widehat{O}_{1}({\bf t_2},n)\cdot \lvacn {\gamma_{+}({\bf  t_{2}})}  \, {\gamma_{+}({\bf  t_{1}})} \, O_{0} \, {\gamma_-({\bf  t_{0}})}\rvacn\\
  &=\widehat{O}_{1}({\bf t_2},n)\cdot  \tau_n^{0,\emptyset}({\bf t_{1}+t_{2}},{\bf t_0}),
\end{split}
\end{equation}
where $\tau_n^{0,\emptyset}({\bf t_{2}},{\bf t_0})= \lvacn {\gamma_{+}({\bf  t_{2}})} \, O_{0} \, {\gamma_-({\bf  t_{0}})}\rvacn$ and
\begin{equation}
\begin{split}\label{O2}
 \tau_n^{1,+}&=\lvacn {\gamma_{+}({\bf  t_{2}})} \, O_{1} \, {\gamma_{+}({\bf  t_{1}})} \, O_{0} \, {\gamma_-({\bf  t_{0}})}\rvacn\\
 &=\widehat{O}_{0}({\bf t_0},n)\cdot \lvacn {\gamma_{+}({\bf  t_{2}})}  \, O_{1} {\gamma_{+}({\bf  t_{1}})} \, {\gamma_-({\bf  t_{0}})}\rvacn\\
  &=\widehat{O}_{0}({\bf t_0},n)\cdot  \exp\left(\sum_{k=1}^\infty k\, t_{0,k} \, t_{1,k} \right) \tau_n^{0,\emptyset}({\bf t_{2}},{\bf t_0}),
\end{split}
\end{equation}
where  $\tau_n^{0,\emptyset}({\bf t_{2}},{\bf t_0})= \lvacn {\gamma_{+}({\bf  t_{2}})} \, O_{1} \, {\gamma_-({\bf  t_{0}})}\rvacn$.

Applying \eqref{Ohyp} and \eqref{Ohyp1} to \eqref{O1} we have
\be\label{o1}
 \tau_n^{1,+}=\widehat{O}_{1}({\bf t_2},n)\cdot \left(\left.\widehat{O}_0({\bf t_1},n)\cdot \exp\left(\sum_{k=1}^\infty k \,t_{0,k}\, t_{1,k}\right)\right|_{{\bf t_1}={\bf t_1+t_2}}\right)
\ee
and
\be\label{o2}
 \tau_n^{1,+}=\widehat{O}_{1}({\bf t_2},n) \widehat{O}_{0}({\bf t_0},n) \cdot \exp\left(\sum_{k=1}^\infty k \, (t_{1,k}+t_{2,k}) \, t_{0,k}\right).
\ee
Applying \eqref{Ohyp} to \eqref{O2} we again arrive at the last equation. However, applying  \eqref{Ohyp1} to  \eqref{O2} we get a new relation
\be\label{o3}
 \tau_n^{1,+}= \widehat{O}_{0}({\bf t_0},n)\cdot  \left(\exp\left(\sum_{k=1}^\infty k\, t_{0,k} \, t_{1,k} \right) \widehat{O}_0({\bf t_0},n)\cdot \exp\left(\sum_{k=1}^\infty k \, t_{0,k} \, t_{2,k}\right)\right).
\ee

Identification of the generating function of the generalized fully simple maps with a nested hypergeometric tau-function allows us to describe the former by the cut-and-join operators $\widehat{W}$, see Section \ref{S4.4}.
If $O=O_{(p)}$, the nested hypergeometric tau-function \eqref{Zvee}  is given by the action of operators  $\widehat{W}_k$ on an elementary tau-function. Namely, from Proposition \ref{prop4.2} we have
\begin{equation}\label{Q2}
\begin{split}
 \tau_n^{1,+}&=\lvacn {\gamma_+({\bf t_2})}\, O_{(p)}^{-1}\, {\gamma_+(\hbar^{-1}{\bf t_1})}\, O_{(p)} \, {\gamma_-(\hbar^{-1}{\bf t_0})} \rvacn\\
 &=\lvacn {\gamma_+({\bf t_2})}\, \exp\left({\frac{1}{\hbar} \sum_{k=1}^\infty t_{1,k}W^{(p)}_{k}} \right)  \, {\gamma_-(\hbar^{-1}{\bf t_0})} \rvacn\\
 &= \exp\left(\frac{1}{\hbar}{\sum_{k=1}^\infty t_{1,k} \widehat{W}^{(p)}_{k}({\bf t_{2}},n)}\right)
 \cdot \exp\left({\frac{1}{\hbar}\sum_{k=1}^\infty k\, t_{0,k} \, t_{2,k}}\right).
\end{split}
\end{equation}
Comparing with expression  \eqref{hyperm} for the hypergeometric tau-function we see that both tau-functions are described by the same cut-and-join operators, but these operators act on different elementary functions.

For $p=1$ and $n=0$ tau-function \eqref{Q2} can be considered as the generating function of the fully simple hypermaps.
Therefore, the generating functions for fully simple hypermaps and ordinary hypermaps are described by the same cut-and-join operators $ \widehat{W}^{(1)}_{k}$.

In particular, for the case of the fully simple maps originally considered by Borot and Garcia-Failde in \cite{BGF}, that is, for the generating function \eqref{FSS}, we have
\begin{equation}
\begin{split}
Z&=\sum_{\lambda_0,\lambda_1} \frac{r^+_{\lambda_0}(\hbar)}{r^+_{\lambda_1}(\hbar)} s_{\lambda_1}({\bf t_2}) s_{\lambda_0/\lambda_1}({\bf t_1}/\hbar)s_{\lambda_0}(\delta_{k,2}/2\hbar)\\
&=  \exp\left({\frac{1}{\hbar}\sum_{k=1}^\infty t_{1,k} \widehat{W}^{(1)}_{k}({\bf t_{2}},0)}\right)  \cdot \exp\left({\frac{1}{\hbar}t_{2,2}}\right),
\end{split}
\end{equation}
which is a fully simple analog of the expression \eqref{smW} for the ordinary maps.

If we slightly modify the normalization of the variables ${\bf t_0}$ and ${\bf t_2}$ for convenience, then the dual description of the nested hypergeometric tau-function \eqref{Q2}  is given by
\begin{equation}
\begin{split}
 \tau_n^{1,+}&=\lvacn {\gamma_+({\bf t_0})}\, O_{(p)}\, {\gamma_-(\hbar^{-1}{\bf t_1})}\, O_{(p)}^{-1} \, {\gamma_-(\hbar^{-1}{\bf t_2})} \rvacn\\
 &= \lvacn {\gamma_+({\bf t_0})}\, \exp \left( {\frac{1}{\hbar} \sum_{k=1}^\infty t_{1,k}W^{(p)}_{-k}} \right) \, {\gamma_-(\hbar^{-1}{\bf t_2})} \rvacn.
\end{split}
\end{equation}
For this nested hypergeometric tau-function we have an alternative cut-and-join description in terms of $\widehat{W}^{(p)}_{-k}$ operators. Namely, from \eqref{dualW} we have
\begin{equation}
\begin{split}
 \tau_n^{1,+}=  \exp\left({\frac{1}{\hbar}\sum_{k=1}^\infty t_{1,k} \widehat{W}^{(p)}_{-k}({\bf t_{0}},n)}\right)
 \cdot \exp\left({\frac{1}{\hbar}\sum_{k=1}^\infty k\,  t_{0,k} \, t_{2,k}}\right).
\end{split}
\end{equation}
which, for $p=1$ and $n=0$ gives a dual expression for the generating function of the fully simple hypermaps.

We see that the operators $\exp \left(\sum_{k=1}^\infty t_k \widehat{W}^{(p)}_{\pm k}\right)$, which act as 
\begin{equation}
\begin{split}
 \exp\left({\sum_{k=1}^\infty t_{k} \widehat{W}^{(p)}_{k}({\bf \tilde {t} },n)}\right) \cdot s_\lambda({\bf \tilde t})&=\sum_{\lambda_1 \subset \lambda} \frac{r_{\lambda,n}}{r_{\lambda_1,n}} s_{\lambda_1}({\bf \tilde t}) s_{\lambda/\lambda_1}({\bf t}),\\
  \exp\left({\sum_{k=1}^\infty t_{k} \widehat{W}^{(p)}_{-k}({\bf \tilde {t} },n)}\right) \cdot s_\lambda({\bf \tilde t})&=\sum_{\lambda_1 \supset \lambda} \frac{r_{\lambda_1,n}}{r_{\lambda,n}} s_{\lambda_1}({\bf \tilde t}) s_{\lambda_1/\lambda}({\bf t})
 \end{split}
\end{equation}
can be described by a composition of the matrix integrals. For instance,
for $p=1$, $n=0$, and $G=(1+N^{-1}z)$ with $N\in {\mathbb Z}_{>0}$ and $\ell(\lambda) \leq N$  we have
\begin{multline}
 \exp\left({\sum_{k=1}^\infty t_{k} \widehat{W}^{(1)}_{k}({\bf \tilde {t} },0)}\right)\cdot s_\lambda({\bf \tilde t}) \\
 =\int_\mathfrak{U} \left[d U\right]  \int_\mathfrak{C} \left[d \mu_\mathfrak{C}({Z})\right]\, s_\lambda (Z^\dagger)  \exp\left(  \Tr \left(N Z U^\dagger +\sum_{k=1}^\infty  \tilde{t}_{k} { U}^k +t_{k}{Z}^k\right)\right)
\end{multline}
and
\begin{multline}
  \exp\left({\sum_{k=1}^\infty t_{k} \widehat{W}^{(1)}_{-k}({\bf \tilde {t} },0)}\right)  \cdot s_\lambda({\bf \tilde t})\\
   = \int_\mathfrak{C} \left[d \mu_\mathfrak{C}({Z})\right] \int_\mathfrak{U} \left[d U\right]  \, s_\lambda(U^\dagger) \exp\left(  \Tr \left(N Z^\dagger U+ \sum_{k=1}^\infty  \tilde{t}_{k} {Z}^k +t_{k}{Z^\dagger}^k\right)\right).
\end{multline}

Let us also consider the matrix models for another choice of the elementary weights in \eqref{Zvee}. For  $O=O_-(\hbar)$ we get a model
\begin{equation}
\begin{split}\label{noW}
 \tau_0^{1,+}&=\lvac {\gamma_+(\hbar^{-1}{\bf t_2})}\, O_-^{-1}(\hbar)\, {\gamma_+(\hbar^{-1}{\bf t_1})}\, O_-(\hbar) \, {\gamma_-({\bf t_0})} \rvac\\
&=\sum_{\lambda_0,\lambda_1}  \frac{r^-_{\lambda_0}(\hbar)}{r^-_{\lambda_1}(\hbar)} s_{\lambda_1}(\hbar^{-1}{\bf t_2}) s_{\lambda_0/\lambda_1}(\hbar^{-1}{\bf t_1})s_{\lambda_0}({\bf t_0}).
\end{split}
\end{equation}
In this case the general construction of Section \ref{S4.2} yields
\begin{multline}\label{matrinv}
\sum_{\ell(\lambda_0),\ell(\lambda_1)\leq N} \frac{r^-_{\lambda_0}(N^{-1})}{r^-_{\lambda_1}(N^{-1})} s_{\lambda_1}(N{\bf t_2}) s_{\lambda_0/\lambda_1}(N{\bf t_1})s_{\lambda_0}({\bf t_0})\\
=\int_\mathfrak{C} \left[d \mu_\mathfrak{C}({Z})\right] 
\int_{\mathfrak{U}^2} \left[d\mu_\mathfrak{U}{(U)}\right] \frac{\displaystyle{\exp\left(\Tr \sum_{k=1}^\infty N\,  t_{2,k}  Z^k+  N\, t_{1,k}  {U^\dagger}^k + {t}_{0,k}  \tilde{U}^k  \right)}}{\det\left(I\otimes I- Z^\dagger\otimes U^\dagger\right)}.
\end{multline}
Again, similar to matrix model \eqref{Simpfs} this expression can be further simplified if we consider the Miwa parametrization of the variables ${\bf t_0}$, that is, $t_{0,k}=\frac{1}{k}\Tr \Lambda^k$ for a diagonal $N\times N$ matrix $\Lambda$,
\begin{multline}
\sum_{\ell(\lambda_0),\ell(\lambda_1)\leq N} \frac{r^-_{\lambda_0}(N^{-1})}{r^-_{\lambda_1}(N^{-1})} s_{\lambda_1}(N{\bf t_2}) s_{\lambda_0/\lambda_1}(N{\bf t_1})s_{\lambda_0}(\Lambda)\\
=\int_\mathfrak{C} \left[d \mu_\mathfrak{C}({Z})\right] 
\int_{\mathfrak{U}} \left[d U\right] \frac{\displaystyle{\exp\left( N\, \Tr\left( \Lambda U+ \sum_{k=1}^\infty  t_{2,k} Z^k +t_{1,k}  {U^\dagger}^k \right)\right)}}{\det\left(I\otimes I- Z^\dagger\otimes U^\dagger\right)}.
\end{multline}
This nested hypergeometric tau-function provides us with a new example of superintegrability
\begin{multline}
r^+_{\lambda_1}(N^{-1}) \sum_{\ell(\lambda_0)\leq N} r^-_{\lambda_0}(N^{-1}) s_{\lambda_0/\lambda_1}(N{\bf t_1})s_{\lambda_0}(\Lambda)\\
=\int_\mathfrak{C} \left[d \mu_\mathfrak{C}({Z})\right] 
\int_{\mathfrak{U}} \left[d U\right] s_{\lambda_1}(Z)\frac{\displaystyle{\exp\left( N\, \Tr\left( \Lambda U+ \sum_{k=1}^\infty t_{1,k}  {U^\dagger}^k \right)\right)}}{\det\left(I\otimes I- Z^\dagger\otimes U^\dagger\right)}.
\end{multline}
\begin{remark}
The operator $O_-^{-1}(\hbar)\, {\gamma_+(\hbar^{-1}{\bf t_1})}\, O_-(\hbar)= O_{p_1}^{+1} \, {\gamma_+(\hbar^{-1}{\bf t_1})}\, O_{p_1}^{-1} $ with $O_{p_1}=O_+(\hbar)$ in \eqref{noW} is not of the form
$O_{p_1}^{-1} {\gamma_+(\hbar^{-1}{\bf t_1})}\, O_{p_1}^{+1}$, therefore
the nested hypergeometric tau-function \eqref{noW} does not belong to the family \eqref{sss}. Hence, Proposition \ref{prop4.2} is not applicable in this case.
\end{remark}

\subsection{$m=2$}

For $m=2$ there are three inequivalent combinations of the signs $\sigma$ for the nested hypergeometric tau-functions:
\begin{equation}
\begin{split}
\tau_n^{2,+,+}&=\lvacn {\gamma_+({\bf t_{3}})}\, O_{2}\, {\gamma_+({\bf t_{2}})} \, O_{1} \, {\gamma_+({\bf t_{1}})}  \, O_{0} \, {\gamma_-({\bf  t_{0}})} \rvacn\\
&=\prod_{j=0}^2 c_n^{(j)}\sum_{\lambda_0,\lambda_1,\lambda_2} 
s_{\lambda_2}({\bf t_{3}})  \,r^{(2)}_{\lambda_2,n} \,   s_{\lambda_1/\lambda_2}{(\bf t_{2}})  \,r^{(1)}_{\lambda_1,n}  \, s_{\lambda_0/\lambda_1}{(\bf t_{1}}) \,r^{(0)}_{\lambda_0,n} \,   s_{\lambda_0}({\bf t_0}),\\
\tau_n^{2,+,-}&=\lvacn {\gamma_+({\bf t_{3}})}\, O_{2}\, {\gamma_+({\bf t_{2}})} \, O_{1} \, {\gamma_-({\bf t_{1}})}  \, O_{0} \, {\gamma_-({\bf  t_{0}})} \rvacn\\
&=\prod_{j=0}^2 c_n^{(j)}\sum_{\lambda_0,\lambda_1,\lambda_2} 
s_{\lambda_2}({\bf t_{3}}) \,r^{(2)}_{\lambda_2,n} \,  s_{\lambda_1/\lambda_2}{(\bf t_{2}}) \,r^{(1)}_{\lambda_1,n}  \,  s_{\lambda_1/\lambda_0}{(\bf t_{1}}) \,r^{(0)}_{\lambda_0,n} \,  s_{\lambda_0}({\bf t_0})\\
\tau_n^{2,-,+}&=\lvacn {\gamma_+({\bf t_{3}})}\, O_{2}\, {\gamma_-({\bf t_{2}})} \, O_{1} \, {\gamma_+({\bf t_{1}})}  \, O_{0} \, {\gamma_-({\bf  t_{0}})} \rvacn,\\
&=\prod_{j=0}^2 c_n^{(j)}\sum_{\lambda_0,\lambda_1,\lambda_2} 
s_{\lambda_2}({\bf t_{3}})\,r^{(2)}_{\lambda_2,n} \,  s_{\lambda_2/\lambda_1}{(\bf t_{2}}) \,r^{(1)}_{\lambda_1,n}  \, s_{\lambda_0/\lambda_1}{(\bf t_{1}}) \,r^{(0)}_{\lambda_0,n} \, s_{\lambda_0}({\bf t_0}).
\end{split}
\end{equation}
The nested structures of the partitions in the sums for these tau-function are respectively $\lambda_2\subset\lambda_1\subset \lambda_0$, $\lambda_2\subset\lambda_1 \supset \lambda_0$, and $\lambda_2 \supset \lambda_1 \subset \lambda_0$.

If the diagonal group operators $O_0$, $O_1$, and $O_2$ correspond to the weight generating functions of the form \eqref{Gcovered}, these nested hypergeometric tau-functions can be described by the chains of the matrix integrals. For instance, let us consider the tau-function $\tau_n^{2,+,+}$ for a particular choice of the diagonal group elements,
namely $O_0=O_+(v_1\hbar), O_1= O_+(v_2\hbar)\, O_+^{-1}(v_1\hbar), O_2=O_+^{-1}(v_2\hbar)$ for some parameters $v_1$, $v_2$, and $\hbar$.
These operators satisfy $O_2\,O_1\,O_0=1$. Then
\begin{equation}
\begin{split}\label{dvapl}
\tau_n^{2,+,+}&=\lvacn {\gamma_+({\bf t_{3}})}\, O_+^{-1}(v_2\hbar)\, {\gamma_+({\bf t_{2}})} O_+(v_2\hbar) O_+^{-1}(v_1\hbar) \, {\gamma_+({\bf t_{1}})}  \, O_+(v_1\hbar) \, {\gamma_-({\bf  t_{0}})} \rvacn\\
&=\prod_{j=0}^2 c_n^{(j)}\sum_{\lambda_0,\lambda_1,\lambda_2}  \frac{r^{+}_{\lambda_0,n}(\hbar v_1)}{r^{+}_{\lambda_1,n}(\hbar v_1)} \frac{r^{+}_{\lambda_1,n}(\hbar v_2)}{r^{+}_{\lambda_2,n}(\hbar v_2)}
s_{\lambda_2}({\bf t_{3}})  \,   s_{\lambda_1/\lambda_2}{(\bf t_{2}}) \, s_{\lambda_0/\lambda_1}{(\bf t_{1}})    s_{\lambda_0}({\bf t_0}).
\end{split}
\end{equation}

Construction of Section \ref{S4.2} immediately leads to a matrix model description of this nested hypergeometric tau-function for $N_1^{-1}=\hbar v_1$, $N_2^{-1}=\hbar v_2$. Since the parameters of several consecutive diagonal group elements coincide, they can be described by the matrix integrals of the same size, which simplifies the chain of matrix integrals. Namely, the resulting matrix model is a chain of two integrals of the form \eqref{Simpfs}:
\begin{equation}
\begin{split}
&\sum_{\ell(\lambda_0),\ell(\lambda_1),\ell(\lambda_2)\leq \min (N_1,N_2)}  \frac{r^{+}_{\lambda_0}(N_1^{-1})}{r^{+}_{\lambda_1}(N_1^{-1})} \frac{r^{+}_{\lambda_1}(N_2^{-1})}{r^{+}_{\lambda_2}(N_2^{-1})}
s_{\lambda_2}({\bf t_{3}})  \,   s_{\lambda_1/\lambda_2}{(\bf t_{2}}) \, s_{\lambda_0/\lambda_1}{(\bf t_{1}})    s_{\lambda_0}({\bf t_0})\\
&=\int_\mathfrak{U} \left[d U_2\right]  \int_\mathfrak{C} \left[d \mu_\mathfrak{C}({Z_2})\right] \int_\mathfrak{U} \left[d U_1\right]  \int_\mathfrak{C} \left[d \mu_\mathfrak{C}({Z_1})\right] \times \\
&\times \frac{\displaystyle{\exp\left( \Tr\left( N U_2^\dagger Z_2 + N U_1^\dagger Z_1+\sum_{k=1}^\infty  t_{3,k} U_2^k + t_{2,k} Z_2^k +t_{1,k}Z_1^k + t_{0,k}{Z_1^\dagger}^k  \right)\right)}}{\det\left(I_1\otimes I_2- U_1 \otimes  Z_2^\dagger\right)}.
\end{split}
\end{equation}
It provides an example of superintegrability:
\begin{equation}
\begin{split}
&{r^{-}_{\lambda_2}(N_2^{-1})} \sum_{\ell(\lambda_0),\ell(\lambda_1)\leq \min (N_1,N_2)}  \frac{r^{+}_{\lambda_0}(N_1^{-1})}{r^{+}_{\lambda_1}(N_1^{-1})} r^{+}_{\lambda_1}(N_2^{-1})
  s_{\lambda_1/\lambda_2}{(\bf t_{2}}) \, s_{\lambda_0/\lambda_1}{(\bf t_{1}})    s_{\lambda_0}({\bf t_0})\\
&=\int_\mathfrak{U} \left[d U_2\right]  \int_\mathfrak{C} \left[d \mu_\mathfrak{C}({Z_2})\right] \int_\mathfrak{U} \left[d U_1\right]  \int_\mathfrak{C} \left[d \mu_\mathfrak{C}({Z_1})\right] \times\\
&\times s_{\lambda_2}(U_2) \frac{\displaystyle{\exp\left( \Tr\left( N U_2^\dagger Z_2 + N U_1^\dagger Z_1+\sum_{k=1}^\infty(  t_{2,k} Z_2^k +t_{1,k}Z_1^k +  t_{0,k}{Z_1^\dagger}^k )\right)\right)}}{\det\left(I_1\otimes I_2- U_1 \otimes  Z_2^\dagger\right)}.
\end{split}
\end{equation}
There is also a similar relation for the variables ${\bf t_0}$.

The nested hypergeometric tau-function \eqref{dvapl} belongs to the family, described by Proposition \ref{prop4.2}, therefore there is a cut-and-join description
\begin{equation}
\begin{split}
\tau_n^{2,+,+}&=\lvacn {\gamma_+({\bf t_{3}})}\,\exp \left({\sum_{k=1}^\infty t_{2,k} W^{(1)}_{k} (v_2)}\right) \exp \left({\sum_{k=1}^\infty t_{1,k} W^{(1)}_{k} (v_1)} \, {\gamma_-({\bf  t_{0}})}\right)  \rvacn\\
 &=  \exp\left({\sum_{k=1}^\infty t_{2,k} \widehat{W}^{(1)}_{k} ({\bf t_3},n,v_2)} \right) \exp\left({\sum_{k=1}^\infty t_{1,k} \widehat{W}^{(1)}_{k} ({\bf t_3},n,v_1)} \right) \cdot \exp\left(\sum_{k=1}^\infty k\, t_{0,k} \, t_{3,k}\right),
\end{split}
\end{equation}
a dual description
\begin{equation}
\begin{split}
\tau_n^{2,+,+}&=\lvacn {\gamma_+({\bf t_{0}})} \,\exp \left({\sum_{k=1}^\infty t_{1,k} W^{(1)}_{-k} (v_1)} \right) \,\exp \left({\sum_{k=1}^\infty t_{2,k} W^{(1)}_{-k} (v_2)}\right) \,{\gamma_-({\bf  t_{3}})} \rvacn\\
 &=\exp\left({\sum_{k=1}^\infty t_{1,k} \widehat{W}^{(1)}_{-k} ({\bf t_0},n,v_1)}\right)  \exp \left({\sum_{k=1}^\infty t_{2,k} \widehat{W}^{(1)}_{-k} ({\bf t_0},n,v_2)} \right)  \cdot \exp \left( \sum_{k=1}^\infty k \, t_{0,k}\, t_{3,k}\right),
\end{split}
\end{equation}
and a combination
\be\label{mixedW}
\tau_n^{2,+,+}=\exp \left( {\sum_{k=1}^\infty t_{1,k} \widehat{W}^{(1)}_{-k} ({\bf t_0},n,v_1)} \right) \exp \left({\sum_{k=1}^\infty t_{2,k} \widehat{W}^{(1)}_{k} ({\bf t_3},n,v_2)}\right)  \cdot \exp \left({\sum_{k=1}^\infty k \, t_{0,k}\, t_{3,k}}\right).
\ee
Here to distinguish the cut-and-join operators we explicitly indicate the parameters of the corresponding weight generating functions.

\section{Broken integrability}\label{S6}

Constructions of this paper allows us also to describe a huge family of functions, which are neither tau-functions of the KP hierarchy nor tau-functions of 2D Toda lattice hierarchy.
\begin{remark}
These functions, in particular, describe the so-called {\em stuffed} versions of the enumerative geometry invariants \cite{Borot} considered in Section \ref{S5}.
\end{remark}
Namely, one can substitute one or several operators $\gamma_{\sigma_i}({\bf t_i})$ with $m\geq i\geq 0$ in \eqref{taudef}
with arbitrary formal series of only positive or only negative bosonic operators ${J}_k$. Without any loss of generality we can consider only the Schur functions $s_\nu({\bf t})$ which constitute a basis in ${\mathbb C} [{\bf t}]$. For $i=0$ the substitution is described in an elementary way by the series expansion of $\gamma_-$, therefore let us consider $m \geq 1$.

The skew Schur function expansion of the obtained vacuum expectation value can be derived in the same way as \eqref{MTS}. Namely, if in the definition of vacuum expectation value \eqref{taudef} we substitute $\gamma_{\sigma_i}({\bf t_i})$  with $s_\nu({\bf \tilde J_\pm})$, where $\tilde J_k = k  J_k$, then in its expansion \eqref{MTS} the corresponding skew Schur function $  s_{(\lambda_{i-1}/\lambda_i)^{\sigma_i}}({\bf t_{i}})$ should be substituted with the Littlewood--Richardson coefficient $c^{\lambda_i}_{\lambda_{i-1}\nu }$ for $s_\nu({\bf \tilde J_+})$ and with  $c^{\lambda_{i-1}}_{\lambda_{i}\nu }$ for $s_\nu({\bf \tilde J_-})$.

Indeed, skew Schur functions can be expanded in the basis of the Schur functions with the Littlewood--Richardson coefficients
\be
s_{\mu/\lambda}=\sum_\nu c^{\mu}_{\lambda \nu} s_\nu,
\ee
then from Proposition \ref{Propsk} we have
\begin{equation}
\begin{split}
s_\nu({\bf \tilde J_+}) \left |\lambda , n\right > &=\sum_{\mu} (-1)^{b(\mu/ \lambda )}  c^{\mu}_{\lambda \nu}  \left |\mu , n\rbr, \\
s_\nu({\bf \tilde J_-}) \left |\lambda , n\right > &= \sum_{\mu} (-1)^{b(\lambda/ \mu )}  c^{\lambda}_{\mu \nu}  \left |\mu , n\rbr.
\end{split}
\end{equation}

For instance, for $m=1$ instead of the expansion \eqref{1+} and \eqref{1-} we have
\begin{equation}
\begin{split}\label{}
\lvacn {\gamma_{+}({\bf  t_{2}})} \, O_{1} \, s_\nu ({\bf \tilde J_+}) \, O_{0} \, {\gamma_-({\bf  t_{0}})}\rvacn
& = c_n^{(0)}c_n^{(1)}\sum_{\lambda_0,\lambda_1}s_{\lambda_1}({\bf t_2})\, r^{(1)}_{\lambda_1,n}\, c^{\lambda_1}_{\lambda_0 \nu} \, 
 r^{(0)}_{\lambda_0,n}\,  s_{\lambda_0}({\bf t_0}),\\
 \lvacn {\gamma_{+}({\bf  t_{2}})} \, O_{1} \, s_\nu ({\bf \tilde J_-}) \, O_{0} \, {\gamma_-({\bf  t_{0}})}\rvacn
& = c_n^{(0)}c_n^{(1)}\sum_{\lambda_0,\lambda_1}s_{\lambda_1}({\bf t_2})\, r^{(1)}_{\lambda_1,n}\, c^{\lambda_0}_{\lambda_1 \nu} \, 
 r^{(0)}_{\lambda_0,n}\,  s_{\lambda_0}({\bf t_0}).
\end{split}
\end{equation}
Using the identity $s_{\lambda/\mu}({\bf t})=\sum_\nu c^{\lambda}_{\mu \nu} s_\nu({\bf t})$ from these identities one can recover \eqref{1+} and \eqref{1-}.

For the weight generating functions of the form \eqref{Gcovered} we can describe these functions by the matrix integrals. 
Applying the Schur function expansion in the variables ${\bf t_1}$ of the matrix integrals \eqref{Simpfs} and \eqref{matrinv}  we have
\begin{multline}
\sum_{\ell(\lambda_0),\ell(\lambda_1)\leq N}  \frac{r^+_{\lambda_0}(N^{-1})}{r^+_{\lambda_1}(N^{-1})} s_{\lambda_1}({\bf t_2})  c^{\lambda_0}_{\lambda_1 \nu} s_{\lambda_0}(N{\bf t_0})\\
=\int_\mathfrak{U} \left[d U\right]  \int_\mathfrak{C} \left[d \mu_\mathfrak{C}({Z})\right] s_\nu (Z)\exp\left( \Tr\left(N\, U^\dagger Z + \sum_{k=1}^\infty \left( t_{2,k} \Tr{U}^k+ N\, t_{0,k}{Z^\dagger}^k\right) \right)\right)
\end{multline}
and
\begin{multline}
\sum_{\ell(\lambda_0),\ell(\lambda_1)\leq N} \frac{r^-_{\lambda_0}(N^{-1})}{r^-_{\lambda_1}(N^{-1})} s_{\lambda_1}(N{\bf t_2})\,  c^{\lambda_0}_{\lambda_1 \nu} \, s_{\lambda_0}({\bf t_0})\\
=\int_\mathfrak{C} \left[d \mu_\mathfrak{C}({Z})\right] 
\int_{\mathfrak{U}^2} \left[d\mu_\mathfrak{U}{(U)}\right] s_\nu(U^\dagger)\,  \frac{\displaystyle{\exp\left(\Tr \sum_{k=1}^\infty N\,  t_{2,k} Z^k  + {t}_{0,k}  \tilde{U}^k  \right)}}{\det\left(I\otimes I- Z^\dagger\otimes U^\dagger\right)}.
\end{multline}
We see that insertion of the arbitrary function of the bosonic operators $J_k$ with a definite sign on the matrix model side corresponds to insertion of the same function of the corresponding matrix argument.
Therefore, if for a nested hypergeometric tau-function with the diagonal group elements described by \eqref{Gcovered} we substitute some of the $\gamma_{\sigma_i}({\bf t_i})$ for $m\geq i\geq 0$ by operators $s_\nu({\bf \tilde J_\pm})$, then a modification of the construction of Section \ref{S4.2} will immediately give us a multi-matrix model with multi-trace potential and, possibly, interaction.

To describe the operators $f({\bf J_\pm })$ one can also use the cut-and-join operators.
For instance, one can consider consider the following combinations of the operators:
\begin{equation}
\begin{split}
O_{(p)} \, f({\bf J_+})  \,O_{(p)}^{-1} &= f({\bf W^{(p)}_+}),\\
O_{(p)}^{-1} \, f({\bf J_-})  \,O_{(p)} &=  f({\bf W^{(p)}_-}) 
\end{split}
\end{equation}
with a diagonal group element $O_{(p)}=\prod_{j=1}^p O_+(u_j)$ and arbitrary $f({\bf t})\in{\mathbb C}[\![{\bf t}]\!]$. 
Then the relations \eqref{Woper} and \eqref{Woper-} can be generalized as follows:
\begin{equation}
\begin{split}
\lvacn \gamma_+({\bf t}) O_{(p)} \, f({\bf J_+})  \,O_{(p)}^{-1}
& =  f({ \widehat{\bf {W}}^{\bf (p)}_{\bf +}})  \cdot \lvacn \gamma_+({\bf t}),\\
\lvacn \gamma_+({\bf t}) O_{(p)}^{-1} \, f({\bf J_-})  \,O_{(p)}
& =  f({ \widehat{\bf {W}}^{\bf (p)}_{\bf -}})  \cdot \lvacn \gamma_+({\bf t}).
\end{split}
\end{equation}

For instance, for the stuffed analogs of the generating functions for the fully simple maps \eqref{Zvee} we have
\begin{equation}
\begin{split}
&\lvacn {\gamma_+({\bf t_2})} \,O_{(p)} \, f_1({\bf J_+})  \,O_{(p)}^{-1} \, f_0({\bf J_-}) \rvacn\\
&=\lvacn {\gamma_+({\bf t_2})} \, f_1({\bf W^{(p)}_{+}}) \,  f_0({\bf J_-})   \rvacn\\
&=f_1({\widehat{\bf W}^{\bf(p)}_{\bf +}}({\bf t_2},n)) \cdot \lvacn {\gamma_+({\bf t_2})} \,  f_0({\bf J_-}) \rvacn\\
&=f_1({\widehat{\bf W}^{\bf(p)}_{\bf +}}({\bf t_2},n)) \cdot f_0(k t_{2,k}).
\end{split}
\end{equation}

\bibliographystyle{alphaurl}
\bibliography{KPTRref}
\end{document}